\documentclass[useAMS,usenatbib,usegraphicx]{mn2e}
\usepackage{bm}
\usepackage{psfig}
\usepackage{amssymb}
\usepackage{amsmath}
\usepackage{multirow}
\usepackage{twoopt}
\usepackage{times}
\usepackage{color}
\usepackage{units}
\usepackage{subfigure}
\usepackage{version}

\usepackage[pdftex,             
            breaklinks=true,%
            colorlinks=true,%
            citecolor=black,
            pdfauthor={Bonnivard et al.},%
            pdftitle={Template for manuscripts MNRAS}%
           ]{hyperref}

\makeatletter

\newcommand{\beq}{\begin{equation}}
\newcommand{\eeq}{\end{equation}}
\newcommandtwoopt{\citeads}[3][][]{\href{http://adsabs.harvard.edu/abs/#3}{\def\hyper@linkstart##1##2{}\let\hyper@linkend\@empty\citealp[#1][#2]{#3}}}
\newcommandtwoopt{\citepads}[3][][]{\href{http://adsabs.harvard.edu/abs/#3}{\def\hyper@linkstart##1##2{}\let\hyper@linkend\@empty\citep[#1][#2]{#3}}}
\newcommandtwoopt{\citetads}[3][][]{\href{http://adsabs.harvard.edu/abs/#3}{\def\hyper@linkstart##1##2{}\let\hyper@linkend\@empty\citet[#1][#2]{#3}}}
\newcommandtwoopt{\citealpads}[3][][]{\href{http://adsabs.harvard.edu/abs/#3}{\def\hyper@linkstart##1##2{}\let\hyper@linkend\@empty\citealp[#1][#2]{#3}}}
\newcommandtwoopt{\citealtads}[3][][]{\href{http://adsabs.harvard.edu/abs/#3}{\def\hyper@linkstart##1##2{}\let\hyper@linkend\@empty\citealt[#1][#2]{#3}}}
\newcommandtwoopt{\citeyearads}[3][][]{\href{http://adsabs.harvard.edu/abs/#3}{\def\hyper@linkstart##1##2{}\let\hyper@linkend\@empty\citeyear[#1][#2]{#3}}}
\newcommandtwoopt{\citeadsstar}[3][][]{\href{http://adsabs.harvard.edu/abs/#3}{\def\hyper@linkstart##1##2{}\let\hyper@linkend\@empty\citealp*[#1][#2]{#3}}}
\newcommandtwoopt{\citepadsstar}[3][][]{\href{http://adsabs.harvard.edu/abs/#3}{\def\hyper@linkstart##1##2{}\let\hyper@linkend\@empty\citep*[#1][#2]{#3}}}
\newcommandtwoopt{\citetadsstar}[3][][]{\href{http://adsabs.harvard.edu/abs/#3}{\def\hyper@linkstart##1##2{}\let\hyper@linkend\@empty\citet*[#1][#2]{#3}}}
\newcommandtwoopt{\citeyearadsstar}[3][][]{\href{http://adsabs.harvard.edu/abs/#3}{\def\hyper@linkstart##1##2{}\let\hyper@linkend\@empty\citeyear*[#1][#2]{#3}}}
\newcommandtwoopt{\citeauthoradsstar}[3][][]{\href{http://adsabs.harvard.edu/abs/#3}{\def\hyper@linkstart##1##2{}\let\hyper@linkend\@empty\citeauthor*[#1][#2]{#3}}}
\newcommandtwoopt{\citepthesis}[3][][]{\href{http://tel.archives-ouvertes.fr/docs/#3}{\def\hyper@linkstart##1##2{}\let\hyper@linkend\@empty\citep[#1][#2]{#3}}}
\newcommandtwoopt{\citetthesis}[3][][]{\href{http://tel.archives-ouvertes.fr/docs/#3}{\def\hyper@linkstart##1##2{}\let\hyper@linkend\@empty\citet[#1][#2]{#3}}}
\makeatother
\begin{document}
\title[Dark matter annihilation and decay in dSphs]{Dark matter annihilation and decay in dwarf spheroidal galaxies: \\ The classical and ultrafaint dSphs} 
\author[Bonnivard, Combet, Daniel et al.]{V. Bonnivard$^{1}$\thanks{E-mails:bonnivard@lpsc.in2p3.fr (VB), dmaurin@lpsc.in2p3.fr (DM), mgwalker@andrew.cmu.edu (MGW)}, 
C. Combet$^{1}$, M. Daniel$^{2}$, S. Funk$^{3}$, A. Geringer-Sameth$^{4,5}$, J.A. Hinton$^{6}$,
\newauthor D. Maurin$^{1}$\footnotemark[1], J. I. Read$^{7}$, S. Sarkar$^{8}$, M. G. Walker$^{4,5}$\footnotemark[1], M. I. Wilkinson$^{9}$ \\
  $^1$LPSC, Universit\'e Grenoble-Alpes, CNRS/IN2P3,
     53 avenue des Martyrs, 38026 Grenoble, France\\
  $^2$Department of Physics, University of Liverpool, Liverpool, L69 7ZE. UK\\
  $^3$Erlangen Center for Astroparticle Physics (ECAP), Friedrich-Alexander Universit\"{a}t Erlangen-N\"{u}rnberg\\
  $^4$Department of Physics, Carnegie Mellon University, Pittsburgh, PA 15213, USA\\
  $^5$McWilliams Center for Cosmology, 5000 Forbes Avenue Pittsburgh, PA 15213, USA\\
  $^6$MPI f\"{u}r Kernphysik, D 69029 Heidelberg, Postf. 103980, Germany\\
  $^7$Department of Physics, University of Surrey, Guildford, GU2 7XH, Surrey, UK\\
  $^{8}$Rudolf Peierls Centre for Theoretical Physics, University of Oxford OX1 3NP, UK \& 
              Niels Bohr Institute, Blegdamsvej 17, Copenhagen 2100, Denmark \\
  $^{9}$Department of Physics and Astronomy, University of Leicester, University Road, Leicester, LE1 7RH, United Kingdom\\
}

\pagerange{\pageref{firstpage}--\pageref{lastpage}} \pubyear{Xxxx}
\date{Accepted Xxxx. Received Xxxx; in original form Xxxx}
\label{firstpage}

\maketitle

\begin{abstract}
Dwarf spheroidal (dSph) galaxies are prime targets for present and
future $\gamma$-ray telescopes hunting for indirect signals of
particle dark matter. The interpretation of the data requires careful
assessment of their dark matter content in order to derive robust
constraints on candidate relic particles. Here, we use an {\it
  optimised} spherical Jeans analysis to reconstruct the
`astrophysical factor' for both annihilating and decaying dark matter
in 21 known dSphs. Improvements with respect to previous works are: (i)
the use of more flexible luminosity and anisotropy profiles to
minimise biases, (ii) the use of weak priors tailored on extensive
sets of contamination-free mock data to improve the confidence
intervals, (iii) systematic cross-checks of binned and unbinned
analyses on mock and real data, and (iv) the use of mock data
including stellar contamination to test the impact on reconstructed
signals. Our analysis provides updated values for the dark matter
content of 8 `classical' and 13 `ultrafaint' dSphs, with the quoted
uncertainties directly linked to the sample size; the more flexible
parametrisation we use results in changes compared to previous
calculations. This translates into our ranking of potentially-brightest
and most robust targets---viz., Ursa Minor, Draco, Sculptor---, and of the more
promising, but uncertain targets---viz., Ursa Major~2, Coma---for
annihilating dark matter. Our analysis of Segue 1 is extremely sensitive
to whether we include or exclude a few marginal member stars, making
this target one of the most uncertain. Our analysis illustrates challenges
that will need to be addressed when inferring the dark matter content of
new `ultrafaint' satellites that are beginning to be discovered in southern
sky surveys.
\end{abstract}


\section{Introduction}
\label{sec:intro}
Constraining particle candidates for dark matter (DM) through indirect
searches for their annihilations or decays has a long history
(\citealtads{1978ApJ...223.1015G}; \citealtads{1978ApJ...223.1032S};
for a recent review see \citealtads{2012AnP...524..479B}). Several
astrophysical messengers (radio/infrared; X-rays; $\gamma$-rays and
neutrinos) and targets (e.g., Galactic centre, dwarf spheroidal
galaxies, galaxy clusters) have been used for such studies
\citepads[e.g.][]{2012PDU.....1..136B,2012Prama..79.1021C,2012PDU.....1....1H}
with inconclusive results so far.

Nearby dwarf spheroidal (dSph) galaxies are particularly promising
targets because of their proximity and low intrinsic astrophysical
backgrounds
\citepads{1990Natur.346...39L,2004PhRvD..69l3501E}. Indeed, several
upper limits have been set on the thermally-averaged self-annihilation
cross section $\langle\sigma v\rangle$ of constituent dark matter
particles using $\gamma$-ray observations of dSphs. The H.E.S.S.
collaboration \citepads{2014PhRvD..90k2012A} in their most recent analysis which
combines 5 dSph galaxies require $\langle \sigma v \rangle < 3.9\times
10^{-24}$\;cm$^3$\;s$^{-1}$ (95\% CL) for ${\cal O}$(TeV) mass
particles. By observing Segue~I for 160 hours, the MAGIC collaboration
infers $\langle \sigma v \rangle \lesssim 1.2\times
10^{-24}$\;cm$^3$\;s$^{-1}$ for $\sim500$\;GeV mass DM annihilating
into the $\tau^+\tau^-$ channel
\citepads{2011arXiv1110.6775P,2014JCAP...02..008A}.\footnote{The
  annihilation channel determines the energy spectrum of the
  $\gamma$-rays, hence the instrumental response.} These values are to
date the strongest limits set by ground-based imaging air Cherenkov
telescopes. At lower energy, observations of several dSphs by the
Fermi-LAT satellite \citepads{2014arXiv1410.2242G,2014PhRvD..89d2001A,2015arXiv150302641F}
provide the most stringent constraint to date with $\langle \sigma v
\rangle < 3\times 10^{-26}$\;cm$^3$\;s$^{-1}$ (95\% CL) for dark
matter particles with a mass below 100 GeV annihilating into the
$b\bar b$ channel. As this is of the order of the thermally-averaged
annihilation cross section required for a weakly interacting massive
particle to make up the dark matter, it has become imperative to
scrutinise these constraints closely.
Intriguingly, \citetads{2015arXiv150302320G} report evidence for $\gamma$-ray
emission from the recently discovered Milky Way satellite Reticulum
II~\citepads{2015arXiv150302079K,2015arXiv150302584T} consistent with
annihilating DM of mass ${\cal O}(10-100)$~GeV scale. The subsequent
analysis of \citetads{2015arXiv150306209H} has confirmed the apparent $\gamma$-ray
excess while \citetads{2015arXiv150302632T} claim no statistically significant
detection. An understanding of Reticulum II's DM content will be
crucial for determining whether this interpretation of the $\gamma$-ray signal
is compatible with non-detections in other dSphs.

In the X-ray band, an unidentified excess at 3.55\;keV detected by
stacking\footnote{Stacking signals from different galaxy clusters
  is an effective approach for decaying dark matter
  \citepads{2012PhRvD..85f3517C} but less so for an annihilating
  particle \citepads{2012MNRAS.425..477N,2012A&A...547A..16M}. }
XMM-Newton galaxy cluster spectra \citepads{2014ApJ...789...13B}, and
by using deep observations of M31 and the Perseus cluster
\citepads{2014PhRvL.113y1301B} has been interpreted as a possible
signature of decaying dark matter. However, this interpretation has
been disputed
\citepads{2014PhLB..733..217P,2014arXiv1408.1699J,2014arXiv1408.4388B,2014arXiv1409.4143B}.
Notably, \citetads{2014PhRvD..90j3506M} find no such excess when
performing a similar analysis on dSphs, possibly ruling out the
decaying DM interpretation.\footnote{This non-detection has motivated
  new models for DM attempting to reconcile both claims,
  e.g. \citetads{2014PhRvD..90l3537C}.}
  
As mentioned, dSph galaxies have been highlighted as targets because
of their high mass-to-luminosity ratio and absence of
astrophysical emission at X-ray and $\gamma$-ray energies
\citepads{1990Natur.346...39L,2004PhRvD..69l3501E}. However to
robustly use these objects to constrain particle DM candidates
requires reliable estimates of the astrophysical factor for a potential
signal (the so-called $J$- and $D$-factors, for annihilation and decay
respectively). This requires careful and optimal evaluation of the DM
distribution in these objects. While cosmological priors or assumed DM
profiles are often used to this end
\citepads{2007PhRvD..76l3509S,2009MNRAS.399.2033P,2013arXiv1309.2641M}, data-driven
approaches have been developed, yielding more reliable results
\citepads{2007PhRvD..75h3526S,2009PhRvD..80b3506E,2015ApJ...801...74G}.

We used such a data-driven approach earlier
\citepads{2011MNRAS.418.1526C} to obtain conservative estimates of the
astrophysical $J$-factors of the 8 `classical' dSphs --- those with
the best-measured stellar kinematics to date. Here, we revisit this
question using the optimised kinematic data analysis developed by
\citetads{2015MNRAS.446.3002B} to obtain reliable values for both $J$-
and $D$-factors for these 8 `classical' dSphs. Additionally, we
include 13 `ultrafaint' dSph galaxies in our analysis. Despite their
sparse and often very uncertain kinematic data, the close proximity of
some of these dSphs (Segue~I is located at only $\sim 23$\;kpc) has
focused attention on them, and the most stringent limits set on $\langle \sigma
v \rangle$ have used such objects. This makes it vital to obtain a
proper measure of both statistical and systematic errors for the astrophysical
factors $J$ and $D$ for these
dSphs (a recent case in point is the tiny Hercules galaxy whose mass
estimate dropped by a factor $\sim 3$ once contaminating foreground
stars were properly weeded out;
\citealtads{2009ApJ...706L.150A}). Especially for pointed observations
where many hours of integration may be lavished on a single target, it
is vital to know not only which dSphs are the most promising targets,
but also which are the most {\it robust} targets. This is our key
goal: to obtain unbiased $J$ and $D$ measures with robust
uncertainties.

The nearby dSphs with good data favour dark matter cores over the
cusps predicted by pure cold dark matter cosmological simulations
\citepads[e.g.][]{2003ApJ...588L..21K,2006MNRAS.368.1073G,2008ApJ...681L..13B,2011ApJ...742...20W,2012MNRAS.426..601C,2012ApJ...754L..39A,2013MNRAS.429L..89A}, 
though this has been disputed at least for some
dSphs \citepads[e.g.][]{2014ApJ...791L...3B,2014MNRAS.441.1584R,2014arXiv1406.6079S}.
Such cores can result from stellar feedback
from galaxy formation. This can be surprisingly effective even for low
stellar mass systems \citepads[e.g.][]{1996MNRAS.283L..72N,2005MNRAS.356..107R,2012MNRAS.421.3464P,2013MNRAS.429.3068T,2014ApJ...789L..17M,2015arXiv150202036O}. Given such theoretical and observational
uncertainties, in this paper we allow for considerable freedom in the
central dark matter density and slope.

This paper is organised as follows. In Sect.~\ref{sec:method}, we
present the Jeans analysis and the input parametrisations for the DM,
light, and anisotropy profiles. We discuss the several likelihood
functions tested to match the data and introduce the astrophysical
factors $J$ and $D$ calculated from the reconstructed DM
profiles. Sect.~\ref{sec:Data} describes the surface brightness and
kinematic data used in the analysis. In Sect.~\ref{sec:setup}, we
present the setup used in the Markov Chain Monte Carlo (MCMC) analysis, in particular our choice
for the priors of the free parameters. We also discuss the choice to
be made for the DM halo size and the possible contamination of the
data. Results are presented in Sect.~\ref{sec:Results}, focusing on
the overall contrast of the signals from dSphs with respect to that
from the Galactic DM halo, and on the ranking of $J$- and $D$-factors,
in comparison to previous works. Conclusions are presented in
Sect.~\ref{sec:conclusions}, while appendices \ref{app:unbinned},
\ref{app:rvir}, and \ref{app:pm} provide further details regarding
specific points addressed in this work.

%
\section{Jeans analysis, likelihood functions and astrophysical $J$- and 
$D$-factors}
\label{sec:method}

\subsection{Jeans analysis}
\label{subsec:Jeans}

The mass density profile of the dSphs is the key input for determining
their $J$- and $D$-factors (Section \ref{subsec:J-factor}). Different
techniques have been developed in order to infer the mass profiles
from stellar kinematic data, such as distribution function modelling,
Schwarzschild and `Made-To-Measure' methods, as well as Jeans analysis
(see recent reviews by
\citealtads{2013NewAR..57...52B,2013PhR...531....1S,2013pss5.book.1039W}). In
this work we focus on the latter, using parametric functions as
ingredients of the spherically symmetric Jeans equation. This
technique has already been widely applied to dSphs
\citepads{2007PhRvD..75h3526S,2010PhRvD..82l3503E,2011MNRAS.418.1526C,2015ApJ...801...74G}. Here,
we apply the findings of \citetads{2015MNRAS.446.3002B}, where an
optimised strategy was proposed to mitigate possible biases introduced
by the Jeans modelling.

\subsubsection{Spherical Jeans equation} Dwarf spheroidal galaxies are
considered as collisionless systems described by their phase-space
distribution function, which obeys the collisionless Boltzmann
equation. Assuming steady-state, spherical symmetry and negligible
rotational support, the second-order Jeans equation is obtained by
integrating moments of the phase-space distribution function
\citepads{2008gady.book.....B}:
\begin{equation}
\frac{1}{\nu}\frac{d}{dr}(\nu\bar{v_r^2}) 
+ 2\frac{\beta_\text{ani}(r)\bar{v_r^2}}{r} = -\frac{GM(r)}{r^2},
\label{eq:jeans}
\end{equation}
where $\nu(r)$, $\bar{v_r^2}(r)$, and $\beta_\text{ani}(r)\equiv
1-\bar{v_{\theta}^2}/\bar{v_r^2}$ are the stellar number density,
velocity dispersion, and velocity anisotropy, respectively. Neglecting
the ($<1\%$) contribution of the stellar component, the enclosed mass at 
radius $r$ can be written as
\begin{equation}
M(r) = 4\pi \int_{0}^{r}\rho_{\rm{DM}}(s)s^2\text{d}s,
\label{eq:mass}
\end{equation}
where $\rho_{\rm DM}(r)$ is the DM mass density profile. The solution
to the Jeans equation relates $M(r)$ to $\nu(r)
\bar{v_r^2}(r)$. However, the internal proper motions of stars in
dSphs are \emph{not} resolved, and only line-of-sight projected
observables can be used:
\begin{equation}
  \sigma_\text{p}^2(R)=\frac{2}{\Sigma(R)}\displaystyle
  \int_{R}^{\infty}\biggl (1-\beta_\text{ani}(r)\frac{R^2}{r^2}\biggr
  ) \frac{\nu(r)\, \bar{v_r^2}(r)\,r}{\sqrt{r^2-R^2}}\mathrm{d}r,
  \label{eq:jeansproject}
\end{equation}
with $R$ the projected radius, $\sigma_\text{p}(R)$ the projected
stellar velocity dispersion, and $\Sigma(R)$ the projected light profile
(or surface brightness) given by
\begin{equation}
   \Sigma(R)\!=\!2\! \int_R^{+\infty} \!\!\!\!\frac{\nu(r)\,r\,dr}{\sqrt{r^2-R^2}}.
  \label{eq:lightproject}
\end{equation}
Note that the velocity anisotropy $\beta_\text{ani}(r)$ \emph{cannot}
be measured directly, in contrast to $\sigma_\text{p}(R)$ and
$\Sigma(R)$. In our approach, parametric models for $\beta_\text{ani}(r)$
and $\rho_\text{DM}(r)$ are assumed in order to compute
$\sigma_\text{p}^2(R)$ via equation~(\ref{eq:jeansproject}). We can then
determine the parameters that reproduce best the measured velocity
dispersion $\sigma_{\rm obs}(R)$.

\subsubsection{Choice of parametric functions}   

\hspace{17 pt}
{\it DM density profile:} \rm Following
\citetads{2011MNRAS.418.1526C}, we do not use a strong cosmological
prior (e.g. assume the profile to be cuspy), as this will bias the
derived astrophysical factors. Instead, we fit the model parameters to
data. We adopt the Einasto parametrisation of the DM density profile
\citepads{2006AJ....132.2685M}:
\begin{equation}
           \rho_{\rm DM}^{\rm Einasto}(r)=\rho_{-2}
           \exp\left\{-\frac{2}{\alpha}\left[\left(\frac{r}{r_{-2}}\right)^\alpha
             -1\right]\right\}\;,
        \label{eq:rho_dm_einasto}
        \end{equation}
where the three free parameters are the logarithmic slope $\alpha$,
the scale radius $r_{-2}$ and the normalisation $\rho_{-2}$. \citetads{2015MNRAS.446.3002B}
  find that the choice of parametrisation --- Zhao-Hernquist or
  Einasto --- has negligible impact on the calculated $J$- or $D$-
  factors and their uncertainties. With fewer free parameters, the
  Einasto parametrisation is more optimal in terms of computational
  time.

{\it Velocity anisotropy profile:} \rm We use the \citetads{2007A&A...471..419B} parametrisation to describe the velocity anisotropy profile:
         \begin{equation}
             \beta_\text{ani}^{\rm Baes}(r) =\frac{\beta_0 +
               \beta_\infty (r/r_a)^\eta}{1+(r/r_a)^\eta}\,,
            \label{eq:beta_baes}
         \end{equation}
where the four free parameters are the central anisotropy $\beta_0$,
the anisotropy at large radii $\beta_\infty$, and the sharpness of the
transition $\eta$ at the scale radius $r_a$. This parametrisation was
found to mitigate some of the biases arising in the Jeans analysis
when using less flexible anisotropy functions with fewer free
parameters (e.g., constant, Osipkov-Merrit --- see
\citealtads{2015MNRAS.446.3002B}).
         
{\it Light profile:} We use a generalised Zhao-Hernquist profile
 \citepads{1990ApJ...356..359H,1996MNRAS.278..488Z} for the stellar
 number density:
\begin{equation}
           \nu^{\rm Zhao}(r)=\frac{\nu_s^{\star}}{(r/r_s^{\star})^\gamma
             [1+(r/r_s^{\star})^\alpha]^{(\beta-\gamma)/\alpha}}\;,
           \label{eq:nu_zhao}
        \end{equation}
the five free parameters of which are the normalisation $\nu_s^{\star}$, the
scale radius $r_s^{\star}$, the inner slope $\gamma$, the outer slope $\beta$,
and the transition slope $\alpha$.  Many studies have used less
flexible parametrisations (e.g., King, Plummer, or exponential
profiles), but the use of these can bias the calculated astrophysical
factors \citepads{2015MNRAS.446.3002B}.

\subsection{Likelihood functions}
\label{subsec:Likelihood}

\subsubsection{Binned and unbinned analyses}
Before fitting the actual dSph kinematic data, we tested both a binned
and an unbinned likelihood function on a set of mock data (mimicking
`ultrafaint' and `classical' dSphs, see Appendix
\ref{app:unbinned}). Both methods have been used in the literature,
but to date, no systematic comparison has been undertaken to test the
merits and limits of each approach (binned analyses can be found in
\citealtads{2007PhRvD..75h3526S,2011MNRAS.418.1526C}; unbinned in
\citealtads{2008ApJ...678..614S,2009JCAP...06..014M,2015ApJ...801...74G}). For the binned
analysis, the velocity dispersion profiles $\sigma_{\rm obs}(R)$ are
built from the individual stellar velocities (see Section
\ref{sec:Data}), and the likelihood function we use is:
\begin{equation}
  \mathcal{L}^{\rm bin}=\! \prod_{i=1}^{N_{\rm bins}}
  \!\frac{(2\pi)^{-1/2}}{\Delta{\sigma_i}(R_i)} \exp\biggl
         [\!-\frac{1}{2}\biggl (\!\frac{\sigma_{\rm obs}(R_{i})
             \!-\!\sigma_\text{p}(R_i)}{\Delta {\sigma}_i(R_i)}\biggr
           )^{2}\biggr ],
  \label{eq:likelihood_binned}
\end{equation}
where 
\begin{equation}
\Delta^2 \sigma_i \!\!=\! \Delta^{2} {\sigma_{\rm obs}}(R_i) \!+\!
\left(\frac{1}{2}\left[\sigma_\text{p}(R_i \!+\! \Delta R_i) \!-\!
  \sigma_\text{p}(R_i \!-\! \Delta R_i) \right] \right)^2\!\!\!\!.
\label{eq:sigma_tot}
\end{equation}
The quantity $\Delta {\sigma_{\rm obs}}(R_i)$ is the error on the
velocity dispersion at the radius $R_i$, and $\Delta R_i$ is the
standard deviation of the radii distribution in the $i$-th bin. This
likelihood allows the uncertainties on both
$\sigma_{\rm obs}$ and $R$ for each bin to be taken into account.

For the unbinned analysis, we assume that the distribution of
line-of-sight stellar velocities is Gaussian, centred on the mean
stellar velocity $\bar{v}$. The likelihood function
reads \citepads{2008ApJ...678..614S}:
\begin{equation}
  \mathcal{L}^{\rm unbin}\!=\! \!\!\prod_{i=1}^{N_{\rm stars}}
  \!\!\!\!  \frac{(2\pi)^{-1/2}}{\sqrt{\sigma_\text{p}^2(R_i)\!+\!\Delta_{v_{
          i}}^{2}}} \exp\biggl [\!-\frac{1}{2}\biggl (\!\frac{(v_{\rm
        i}
      \!-\!\bar{v})^{2}}{\sigma_\text{p}^2(R_i)\!+\!\Delta_{v_{i}}^{2}}\biggr
    )\biggr ],
  \label{eq:likelihood_unbinned}
\end{equation}
where the dispersion of velocities at radius $R_i$ of the $i$-th star
comes from both the intrinsic dispersion $\sigma_\text{p}(R_i)$ from
equation~(\ref{eq:jeansproject}) and the measurement uncertainty
$\Delta_{v_{i}}$.

As detailed in Appendix \ref{app:unbinned} the unbinned analysis
reduces the statistical uncertainties on the astrophysical factors,
particularly for the `ultrafaint' dSphs, without introducing
biases. In the remainder of the paper we therefore favour the unbinned
analysis and the binned likelihood is used only to cross-check our
results.

\subsubsection{Analysis with and without membership probabilities}
\label{jfact_wo_pm}
Kinematic samples are often contaminated by interlopers from the Milky
Way (MW) foreground stars. Different methods can be used in order to
remove those contaminants, based e.g. on sigma-clipping, virial
theorem \citepads{2007MNRAS.378..353K,2007MNRAS.377..843W}, or
expectation maximisation (EM) algorithms
\citepads{2009AJ....137.3109W,2009JCAP...06..014M}. The latter allows in particular the
estimation of membership probabilities for each star of the
object. These probabilities can be used as weights when building the
velocity dispersion profile $\sigma_{\rm obs}(R)$ in the binned case,
or used directly in the unbinned likelihood, where
equation~(\ref{eq:likelihood_unbinned}) becomes
\begin{equation}
  \mathcal{L}^{\rm unbin}_{W}\!\!\!=\!\!\! \prod_{i=1}^{\!\!N_{\rm
      stars}}\!\!
  \!\left(\!\frac{(2\pi)^{-1/2}}{\sqrt{\sigma_\text{p}^2(R_i)\!+\!\Delta_{v_{
          i}}^{2}}} \!\exp\!\biggl [\!-\frac{1}{2}\!\biggl
    (\!\frac{(v_{\rm i}
      \!-\!\bar{v})^{2}}{\!\sigma_\text{p}^2(R_i)\!+\!\Delta_{v_{i}}^{2}\!}
    \biggr )\biggr ] \!\right)^{\!P_{i}\!}\!\!\!\!\!,
  \label{eq:likelihood_unbinned_weights}
\end{equation}
with $P_{i}$ the membership probability of the $i$-th star. Another
option when dealing with foreground contamination is to use the
unweighted likelihoods $\mathcal{L}^{\rm bin/unbin}$ but run the
analysis only with stars having large-enough membership probabilities
(typically, $P_i > 0.95$). Whenever membership probabilities are
available, we test both methods (see Section \ref{subsec:pm}).

\subsection{Astrophysical factor for annihilation and decay}
\label{subsec:J-factor}

The $\gamma$-ray differential flux from DM annihilation
(resp. decay) in a dSph galaxy, measured within a solid angle $\Delta
\Omega$, is \citepads{1998APh.....9..137B}
\begin{equation}
\frac{\text{d} \phi_{\gamma}}{\text{d} E_{\gamma}} =
\phi^{\text{PP}}_{J}(E_{\gamma}) \!\times\! J(\Delta \Omega) \quad
\left(\text{resp.~} \phi^{\text{PP}}_{D}(E_{\gamma}) \times
D(\Delta \Omega)\!\right).
\end{equation}
The quantity $\phi^{\text{PP}}_{J}(E_{\gamma})$
$\left(\text{resp.}~\phi^{\text{PP}}_{D}(E_{\gamma})\right)$ is sensitive
to the particle physics, e.g. the annihilation or decay channel. We
focus here on the `astrophysical factor', $J$ (resp. $D$),
 \begin{equation}
      \!\!J \!=\! \!\int\!\!\!\!\!\int\!\! \rho_{\rm DM}^2 (l,\Omega)
      \,dld\Omega \quad\left({\rm \!resp.~} D \!=\!\!
      \int\!\!\!\!\!\int \!\!\rho_{\rm DM}(l,\Omega)
      \,dld\Omega\right),
      \label{eq:J}
 \end{equation} 
which corresponds to the integration along the line-of-sight of the DM
density squared (resp. DM density) over the solid angle $\Delta\Omega
= 2\pi\times[1-\cos(\alpha_{\rm int})]$, $\alpha_{\rm int}$ being the
integration angle. A precise evaluation of this quantity is necessary
for setting robust constraints on the properties of the DM particle,
and is also a useful proxy to rank possible targets according to the
magnitude of their flux. In order to compute the astrophysical factor,
both the density distribution of the DM halo and its extent are
required to be known. Results may be sensitive to the choice of the
latter, as discussed further in Section~\ref{subsec:size} and
Appendix~\ref{app:rvir}.

N-body simulations within the context of $\Lambda$CDM cosmology have
shown that DM halos should contain a large number of smaller
sub-halos. Such sub-structures could significantly increase the
$J$-factors, but the smaller the host halo mass, the less boosted is
the signal (e.g., \citealtads{2014MNRAS.442.2271S}). For DM halos
typical of dSphs, \citetads{2011MNRAS.418.1526C} found no significant
impact of the sub-structures on the $J$-factors so we neglect their
contribution here.

$J$- and $D$-factors of dSphs were found to be best constrained at the
so-called `critical' integration angle $\alpha_c$
\citepads{2011ApJ...733L..46W,2011MNRAS.418.1526C,2015MNRAS.446.3002B}. This
is related to the half-light radius of the dSph, $r_h$, and to its
distance $d$, and differs for $J$- and $D$-factors: $\alpha_{c}^{J}
\sim 2 r_h/d$, while $\alpha_{c}^{D} \sim r_h/d$ (see figure
\ref{fig:J_Fornax} for illustration).

All calculations of astrophysical factors are done with the {\tt
  CLUMPY} code \citepads{2012CoPhC.183..656C}. A new module has been
specifically developed to perform the Jeans analysis, and this upgrade
will be publicly available in the forthcoming second release of the
software (Bonnivard et al., in preparation). The $J$- and $D$-factors
obtained for the eight `classical' and thirteen `ultrafaint' dSph
galaxies are presented in Section~\ref{sec:Results}.

\section{Dwarf spheroidal galaxy data}
\label{sec:Data}

\subsection{Surface brightness data}
\label{subsec:Light}

For each dSph, we estimate $\nu(r)$ by fitting publicly available
photometric data with a Zhao-Hernquist model (equation~\ref{eq:nu_zhao}),
where $r$ is the distance from the dwarf's centre in 3D.  Due to
differences in the nature of the available data for `classical' and
`ultrafaint' dSphs, we adopt different approaches for applying
this model to each category.

For `classical' dSphs, the most homogeneous data sets for estimating
structural parameters remain those of \citetads[][`IH95'
  hereafter]{1995MNRAS.277.1354I}.  IH95 tabulate stellar surface
density profiles in terms of stars counted within concentric
elliptical annuli, with each ellipse (with semi-major and semi-minor
axes $a$ and $b$ respectively) having the same ellipticity and
orientation that IH95 estimate for the dSph as a whole.  IH95 measure
global ellipticities ranging from $e\equiv 1-b/a=0.13$ (Leo II) to
$e=0.56$ (UMi), with a median of $e=0.32$. Because our Jeans models
assumes spherical symmetry, we transform IH95's elliptical annuli into
circular annuli by replacing the `elliptical radii' in their data
tables with geometric means, $R_{\rm gm}\equiv \sqrt{ab}$.

To the circularised, binned surface density profiles, $\Sigma(R)$, we
then fit 2D projections of $\nu(r)$ according to the likelihood
function
\begin{equation}
  \mathcal{L}_{\rm 1}\propto \displaystyle\prod_{i=1}^{N_{\rm bins}}
  \exp\biggl [-\frac{1}{2}\frac{(\Sigma(R_i)- \Sigma_{\rm
        model}(R_i))^2}{\sigma^2_{\Sigma(R_i)}}\biggr ],
  \label{eq:ih95like}
\end{equation}
where $\sigma_{\Sigma(R_i)}$ is the Poisson error associated with the
number of stars counted in the $i^{\rm th}$ bin and the model surface
density is
\begin{equation}
  \Sigma_{\rm model}(R)\equiv
  2\displaystyle\int_{R}^\infty\frac{\nu(r)r}{\sqrt{r^2-R^2}}dr+\Sigma_{\rm
    bkd},
  \label{eq:sbpmodel}
\end{equation}
i.e., the sum of the projection of $\nu(r)$ and a uniform background
density.

For the `ultrafaint' satellites, the largest homogeneous data set is
from the Sloan Digital Sky Survey, which provides positions, colours
and magnitudes of individual stars detected as point sources.  For
each `ultrafaint' satellite, we identify possible members as red giant
branch (RGB) candidates, which we define as point sources whose $g-r$
colours place them within $0.25$ dex of the Dartmouth isochrone
\citepads{2008ApJS..178...89D} calculated for a stellar population
with age 12 Gyr, metallicity corresponding to the mean value estimated
from spectroscopy, and shifted by the distance modulus estimated for
that satellite \citepads{2012AJ....144....4M}.  To the unbinned
distribution of projected positions for $N$ RGB candidates, we fit 2D
projections of $\nu(r)$ according to the likelihood function
\begin{equation}
  \mathcal{L}_2\propto \displaystyle\prod_{i=1}^{N} \Sigma_{\rm model}(R_i),
\end{equation}
where $\Sigma_{\rm model}(R_i)$ is defined as in equation~(\ref{eq:sbpmodel}).

For both `classical' and `ultrafaint' satellites, we adopt uniform
priors on all model parameters and then use the software package
{\tt MultiNest}
\citepads{2008MNRAS.384..449F,2009MNRAS.398.1601F,2013arXiv1306.2144F}
to sample the posterior probability distribution (PDF) of parameter
space.  In a separate contribution, Walker (in preparation) will
discuss detailed results from these fits.  For our present purposes,
we use these samples from the posterior PDFs to propagate
uncertainties in $\nu(r)$ through our estimation of the dark matter
density profiles (see Section \ref{sec:setup}).

\subsection{Kinematic data}
\label{subsec:Kinematics}

We use stellar-kinematic data, in the form of projected positions and
line-of-sight velocities for individual stars, compiled from the
literature.  For all galaxies except Draco, we use the same data sets
as analysed by \citetads{2015ApJ...801...74G} (who provide a detailed
description). For Draco, we adopt the data set of
\citetads{2015arXiv150302589W}, which includes measurements for $\sim
500$ members. 

The kinematic samples for the eight `classical' dSphs include, for each
star, a probability of membership, $P_i$, that is estimated using an
expectation-maximisation algorithm \citepads{2009AJ....137.3109W}.  The
sample for Segue~I \citepads{2011ApJ...733...46S} includes membership
probabilities that are estimated in two ways: via the same EM algorithm,
and alternatively using a Bayesian analysis that considers the entire
data set to be a mixture of Segue~I and foreground populations.  We use
the latter estimation in our analysis -- details of the Segue~I case will be presented in a separate contribution (Bonnivard, Maurin \& Walker, in prep). Data for
the remaining `ultrafaints' do not include membership probabilities, but
rather a binary classification of each star as member or nonmember based
on velocity and line-strength criteria (e.g.,
\citealtads{2007ApJ...670..313S}).  We treat these classifications as
membership probabilities in our analysis, but their values necessarily
are either $P_i=0$ (non-member) or $P_i=1$ (member).

In order to estimate velocity dispersion profiles, we divide each data
set, consisting of $N_{\rm mem}\equiv \Sigma_{i=1}^N P_i$ member stars,
into $\sim \sqrt{N_{\rm mem}}$ bins that each contain a number of stars
whose membership probabilities add to $\sim \sqrt{N_{\rm mem}}$. For each
bin, we estimate velocity dispersion using the maximum-likelihood
procedure described by \citetads{2006AJ....131.2114W}, with membership
probabilities introduced as weights on each star. Our results are based
on the unbinned analysis (see Section \ref{subsec:Likelihood}), and we
use these velocity dispersion profiles only for cross-check purpose. 

\begin{figure*}
\includegraphics[width=\linewidth]{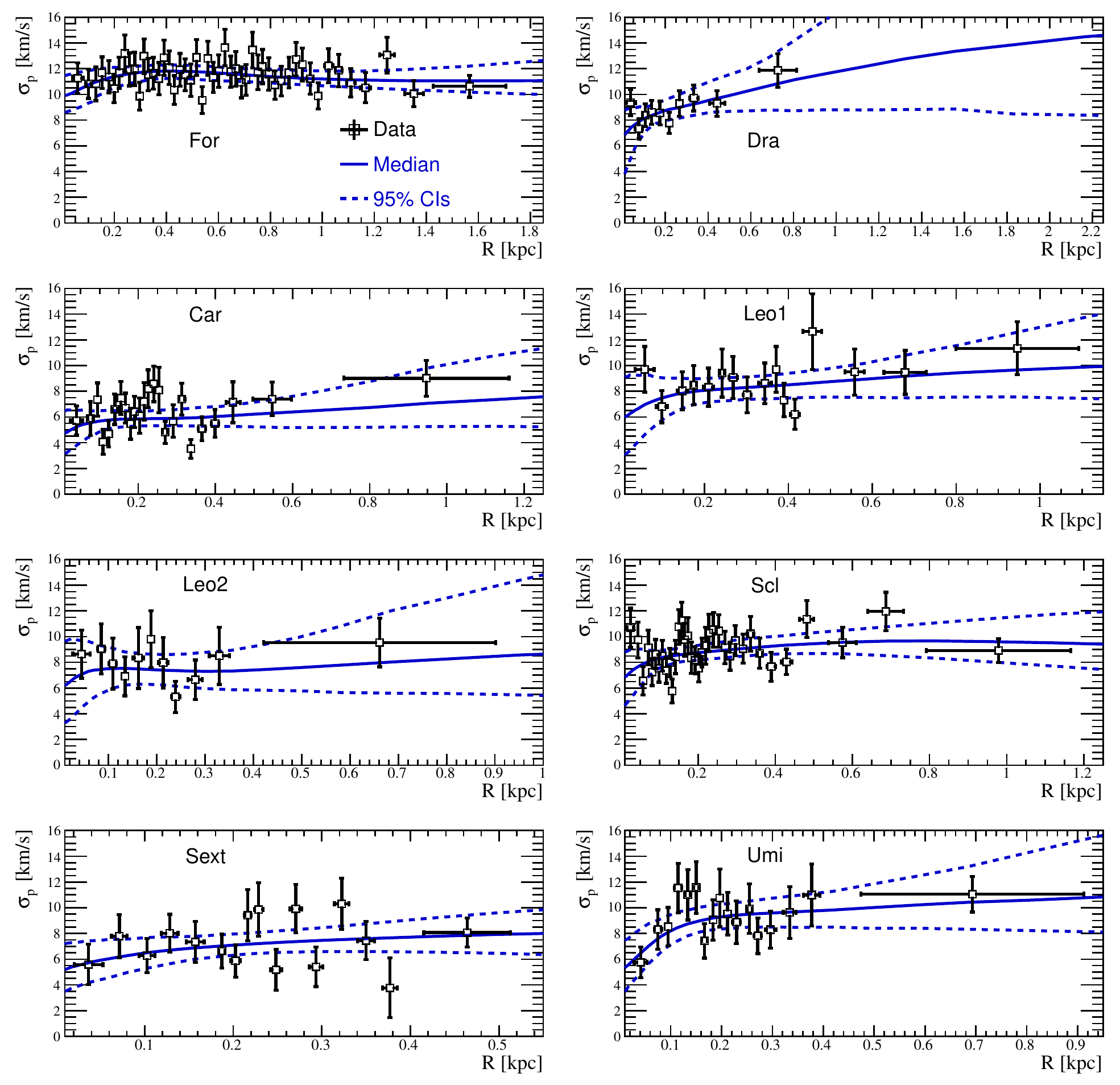}
\caption{Velocity dispersion profiles $\sigma_{p}$ of the eight
  `classical' dSphs: data (symbols) and reconstructed median and 95\%
  CIs (blue lines). These profiles are shown for illustration; our results are based on an unbinned analysis.}
\label{fig:disp_classical}
\end{figure*}
\begin{figure*}
\includegraphics[width=\linewidth]{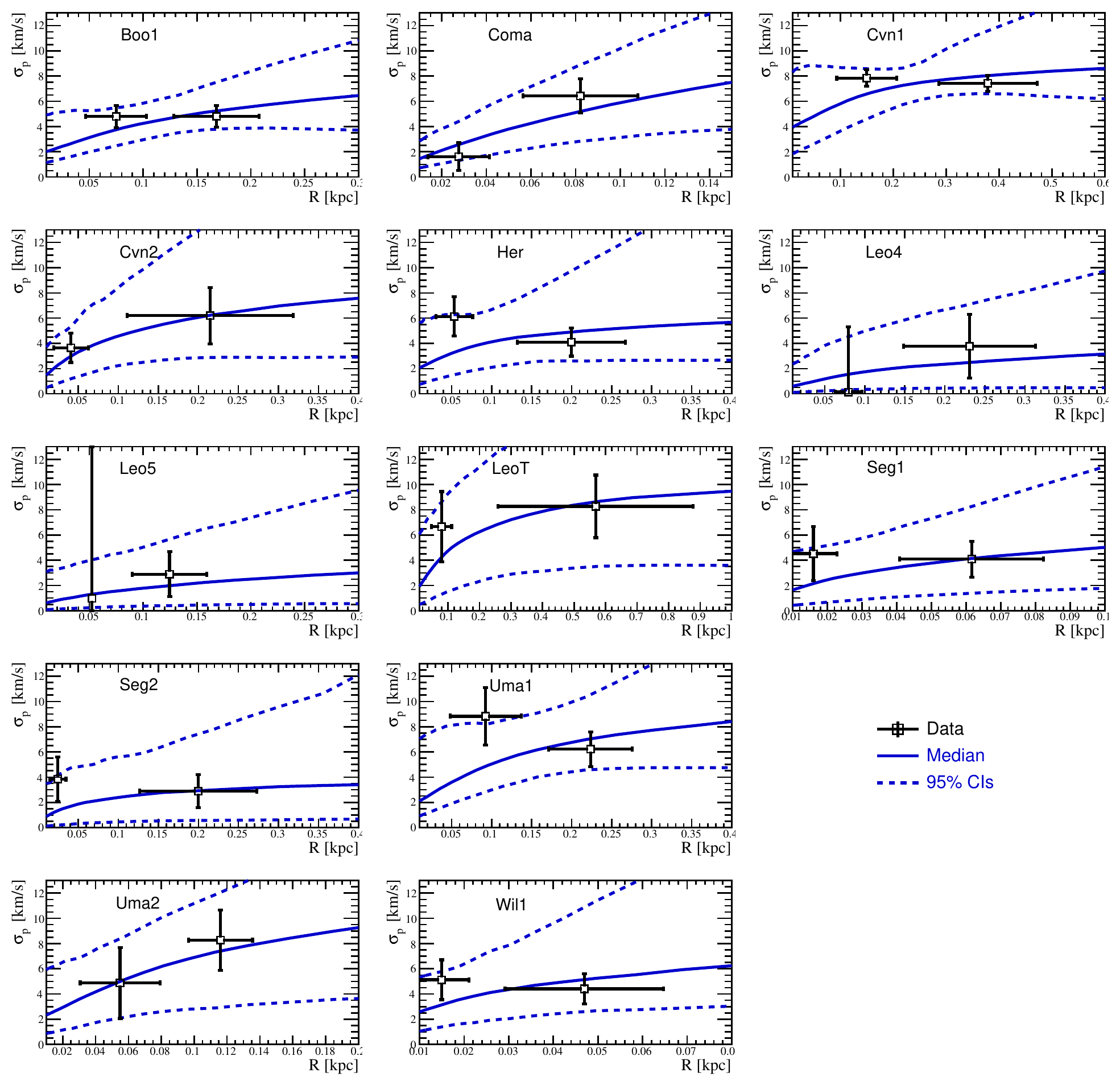}
\caption{Velocity dispersion profiles $\sigma_{p}$ of the thirteen
  `ultrafaint' dSphs: data (symbols) and reconstructed median and 95\%
  CIs (blue lines). These profiles are shown for illustration; our results are based on an unbinned analysis.}
\label{fig:disp_ultrafaint}
\end{figure*}

\section{Analysis setup}
\label{sec:setup}
\subsection{MCMC analysis}
\label{subsec:MCMC}
For each dSph, we perform the Jeans analysis to find the DM
density profile parameters that best fit the stellar kinematics, and
determine their uncertainties. The likelihood functions of the
analysis (Section~\ref{subsec:Likelihood}) have seven free parameters
(see below). To efficiently explore this large parameter space, we use
an MCMC technique, based on Bayesian parameter inference, which allows
us to sample the PDF of a set of free parameters using Markov
chains. To this purpose, we use the Grenoble Analysis Toolkit ({\tt
  GreAT}) \citepads{2011ICRC....6..260P,Putze:2014aba}, which relies
on the Metropolis-Hastings algorithm
\citepads{1953JChPh..21.1087M,hastings70}. The posterior distributions
are obtained after several post-processing steps (burn-in length
removal, correlation length estimation, and thinning of the chains),
allowing a selection of independent samples, insensitive to the
initial conditions. From these PDFs, credibility intervals (CIs) for
any quantity of interest can be easily computed.

For each step of the chains, we randomly select a light profile
parametrisation from the accepted configurations of the MCMC analysis
done previously for the surface brightness data (see Section
\ref{sec:Data}), and use it for computing the velocity dispersion
$\sigma_\text{p}$ (equation \ref{eq:jeansproject}). This effectively propagates
the light profile uncertainties to the posterior distributions of the
DM and anisotropy parameters.

\subsection{Free parameters and priors}
\label{subsec:priors}
The free parameters of the analysis are the three parameters of the
Einasto DM density profile, and the four parameters of the Baes \& van
Hese (2007) velocity anisotropy profile, for which we adopt uniform
priors. We follow \citetads{2015MNRAS.446.3002B} who 
after a thorough study of mock data identified an optimal
prior combination that mitigates several biases introduced by the
Jeans analysis. Table~\ref{table:priors} summarises the ranges used on
each parameter's prior.

\begin{table}
\begin{center}
\caption{Range of uniform priors used for the DM density and velocity
  anisotropy profile parameters.  Note that all models must satisfy
  the \textit{Global Density-Slope Anisotropy Inequality}
  (equation~\ref{eq:slope_aniso}), which reduces the effective range of the
  anisotropy parameters.  See text for details.}
\label{table:priors}
\begin{tabular}{cccc}\hline\hline
 Quantity & Profile & Parameter & Prior range \\
 \hline
DM density \!\!\!\!\!\! &  `Einasto' \!\!\!\!\!\!\!\!\! & $\log_{10}(\rho_{-2}/\text{M}_{\odot}~\text{kpc}^{-3})$\!\!\!\!\!\! & $[5,13]$ \!\!\!\!\!\! \\
\!\!\!\!\!\! & equation~(\ref{eq:rho_dm_einasto}) \!\!\!\!\!\!\!\!\! & $\log_{10}(\text{r}_{-2}/\text{kpc})$\!\!\!\!\!\! & $[\log_{10}(r_{\rm s}^*),1]$ \!\!\!\!\!\! \\\vspace{5mm}
\!\!\!\!\!\! &              \!\!\!\!\!\!\!\!\!    & $\alpha$               \!\!\!\!\!\! & $[0.12,1]$\!\!\!\!\!\!\\ 
Anisotropy \!\!\!\!\!\! & `Baes \& van Hese' \!\!\!\!\!\!\!\!\! & $\beta_{0}$ \!\!\!\!\!\! & $[$-$9,1]$ \!\!\!\!\!\!\\
\!\!\!\!\!\! & equation~(\ref{eq:beta_baes}) \!\!\!\!\!\!\!\!\! & $\beta_{\infty}$  \!\!\!\!\!\! & $[$-$9,1]$ \!\!\!\!\!\!\\
\!\!\!\!\!\! &                              \!\!\!\!\!\!\!\!\! & $\log_{10}(r_{a})$ \!\!\!\!\!\! & $[$-$3,1]$ \!\!\!\!\!\!\\
\!\!\!\!\!\! &                             \!\!\!\!\!\!\!\!\! & $\eta$             \!\!\!\!\!\! & $[0.1,4]$ \!\!\!\!\!\!\\ 
 \hline
\end{tabular}
\end{center}
\end{table}

\it{DM density profile.} \rm \citetads{2015MNRAS.446.3002B} suggested
two cuts on the DM density profile priors in order to tighten the
constraints on the astrophysical factors without introducing bias. First, the
scale radius $r_{-2}$ is forced to be larger than the scale radius of
the stellar component, $r_{\rm s}^{*}$. This drastically reduces the
upper CIs on the astrophysical factors for `ultrafaint' dSphs. A
second cut on the logarithmic slope $\alpha$, $\alpha \geq 0.12$, is
also advocated in this previous work. We have chosen to use both cuts
in our analysis, so as to obtain the most stringent and robust estimates of the
astrophysical factors.

\it{Velocity anisotropy profile.} \rm The priors we use for the 4
parameters of the Baes \& van Hese (2007) anisotropy profiles are also those
of \citetads{2015MNRAS.446.3002B}. The generality of
this parametrisation avoids the bias from the use of
more specific anisotropy profiles, especially for large
data samples. We also implement the \textit{Global Density-Slope
  Anisotropy Inequality} \citepads{2010MNRAS.408.1070C}, which ensures
that solutions to the Jeans equation given by our MCMC analysis
correspond to physical models with positive  phase-space distribution
function:
\begin{equation}
\label{eq:slope_aniso}
\beta_{\text{ani}}(r) \leq
-\frac{1}{2}\frac{\text{d}\log\nu(r)}{\text{d}\log(r)}.
\end{equation}
This reduces the range of allowed velocity anisotropy parameters
depending on the stellar number density $\nu(r)$.

\subsection{Size of the DM halo}
\label{subsec:size}
The extension of the DM halo is needed when computing the $J$- (or
$D$-) factor (equation~\ref{eq:J}). The latter reaches a maximum when
the integration angle $\alpha_{\rm int}$ corresponds to the angular
size of the halo and saturates beyond (see figure \ref{fig:J_Fornax},
where the $J$-factor obtained for Fornax is plotted as a function of
the integration angle $\alpha_{\rm int}$; the median value is seen to
saturate above $\sim 1^\circ$).

There is no clear criterion to define the size of DM halos hence we
adopt two different approaches for each dSph galaxy and for each set
of DM parameters accepted by the MCMC. The first method considers the
tidal radius $r_\text{t}$ to be a good estimator of the halo size (as
shown by N-body or hydrodynamical simulations --- see
e.g. \citealtads{2008MNRAS.391.1685S,2015MNRAS.447.1353M}); this is
computed as:
\begin{equation}
r_\text{t} = \left[ \frac{M_{\rm halo}(r_\text{t})}{[2-\text{d}\ln
      M_{\text{MW}}/\text{d}\ln r (d)]\times M_{\text{MW}}(d)}
  \right]^{(1/3)} \times d\;,
\end{equation}
where $M_{\text{MW}}(d)$ is the mass of the MW enclosed within the
galactocentric distance $d$ of the dSph and $M_{\rm halo}$ is the mass
of the dSph galaxy. A second method to estimate the size of a DM
halo consists in determining the radius $r_{\rm eq}$ where the halo
density is equal to the density of the MW halo, namely,
\begin{equation}
\rho_{\rm DM}^{\rm halo}(r_{\rm eq}) = \rho_{\rm DM}^{\rm MW}(d - r_{\rm eq}).
\end{equation}
We have used both a Navarro-Frenk-White \citepads{1997ApJ...490..493N,2005MNRAS.364..433B}
and an Einasto profile \citepads{2004MNRAS.349.1039N,2008MNRAS.391.1685S}
for the MW density, with no impact on the results. We also find the
above two estimates of the dSph galaxy size to be comparable, leading
to very similar astrophysical factors (see Appendix
\ref{app:rvir}). Note that other assumptions can also be made to
estimate the dSph halo size. For instance, in order to be conservative,
\citetads{2015ApJ...801...74G} used the outermost observed star as
truncation radius for computing the astrophysical factors. However,
this can underestimate the $J$-
and $D$-factors, and may moreover underestimate the credibility
intervals (see Appendix \ref{app:rvir}).

\subsection{Membership probabilities and impact of contamination}
\label{subsec:pm}

$J$- and $D$-factor reconstruction through Jeans analysis has been
previously studied and optimised for contamination-free mock data
\citepads{2011MNRAS.418.1526C,2015MNRAS.446.3002B}. However, actual
observations yield kinematic samples that may be contaminated by field
stars belonging to the Milky Way or to a Galactic stream.  The
conventional approach to handle these interlopers relies on the
arbitrary definition of some threshold separating members from
outliers (sigma-clipping method, see
e.g. \citealtads{1977ApJ...214..347Y}).  Expectation maximisation (EM)
algorithms \citepads{2009AJ....137.3109W} differ from sigma-clipping
methods as they provide membership probabilities $P_i$ for each star
of the sample, which can be used as weights in subsequent analyses,
e.g., equation~(\ref{eq:likelihood_unbinned_weights}).  The EM
algorithm was shown to provide accurate and reliable membership
probabilities in most cases, although some failures may occur for
samples presenting the heaviest contamination and the most overlapping
velocity distributions \citepads{2009AJ....137.3109W}.

\begin{figure}
\includegraphics[width=\columnwidth]{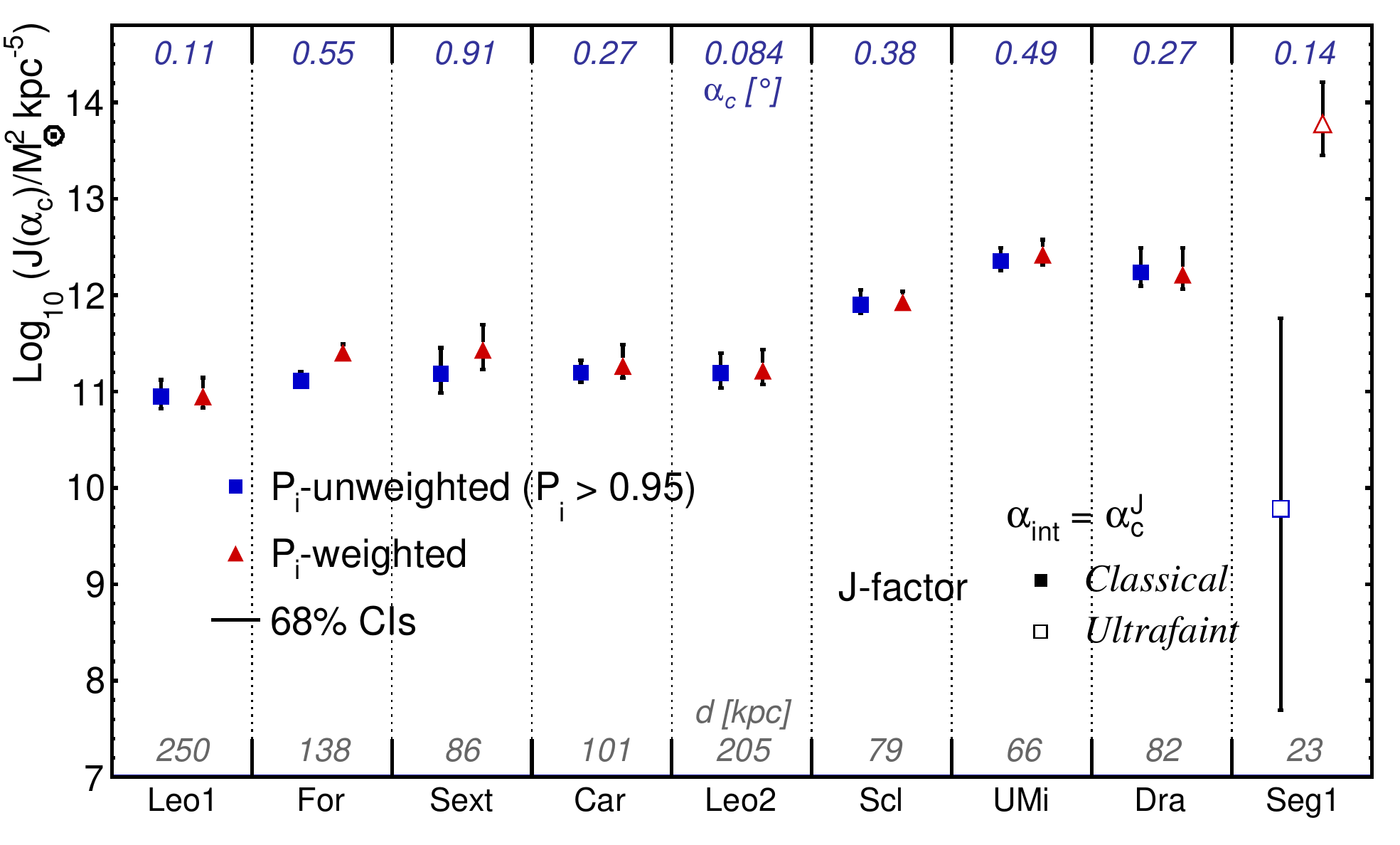}
\includegraphics[width=\columnwidth]{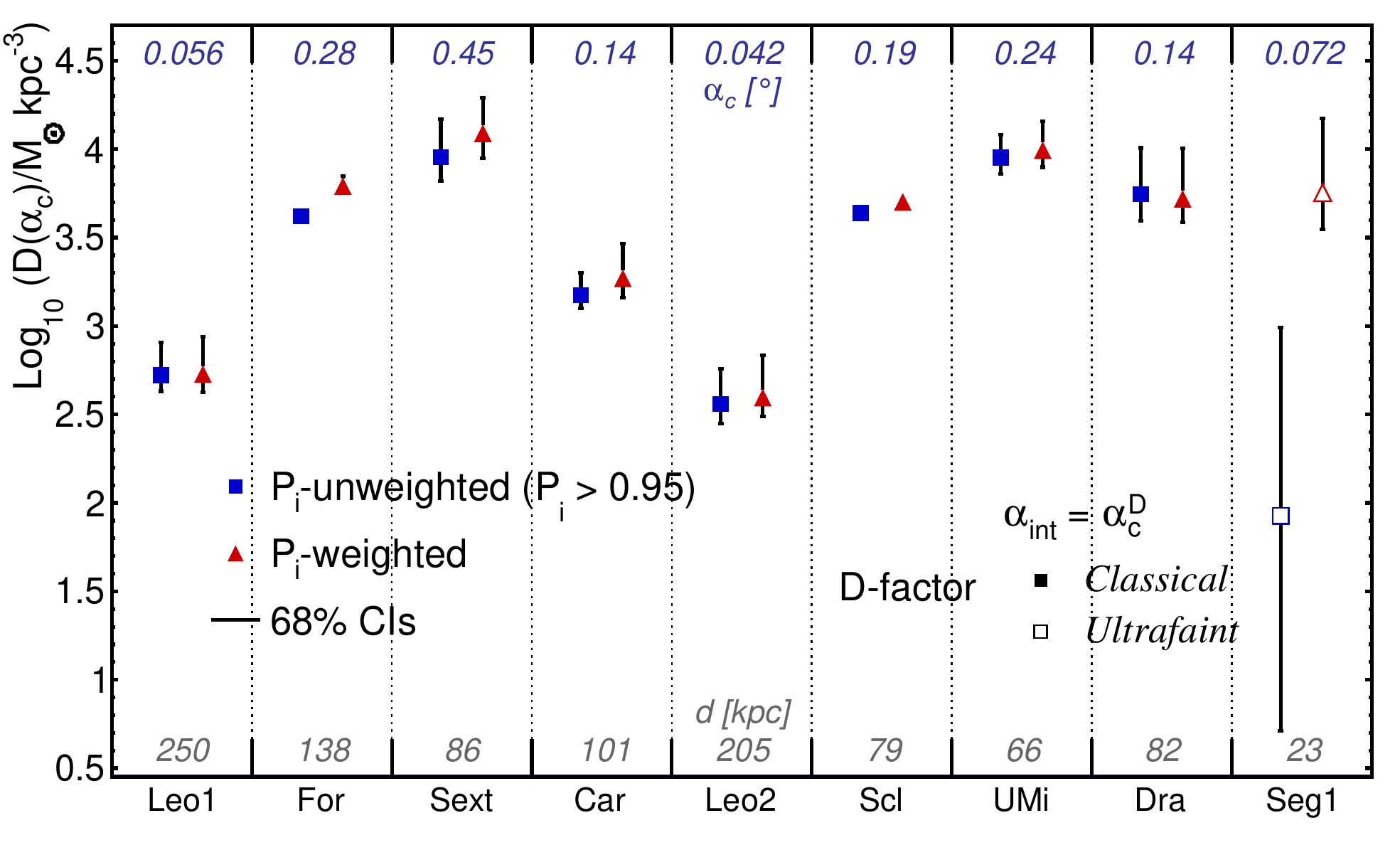}
\caption{$J$- and $D$-factors at the critical integration angle (top
  and bottom panel respectively) for dSph galaxies whose $P_i$ values
  are available. Shown are values calculated using
  equation~(\ref{eq:likelihood_unbinned_weights}), i.e. with
  membership probabilities as weights in the Jeans analysis (red
  triangles), or using all stars passing the cut $P_i>0.95$ (blue
  squares).}
\label{fig:wo_contamination}
\end{figure}

In order to investigate whether residual contamination affects the
$J$-factor values, we either use the star membership probability as
weights in the likelihood function
equation~(\ref{eq:likelihood_unbinned_weights}), or only retain stars
which almost certainly belong to the dSph
($P_i>0.95$). Figure~\ref{fig:wo_contamination} compares these two
approaches for dSphs with $P_i$ values available from the literature
(eight `classical' and Segue~I).  The same $J$-factors ($D$-factors) are
reconstructed in both cases, except for Fornax and Segue~I. For these
two objects, the discrepancy hints at the presence of contamination in
their samples that is not captured by the $P_i$ indicator.
The case of Segue~I will be thoroughly discussed in Bonnivard, Maurin \& Walker (in prep.), and we refer the reader to Appendix~\ref{app:pm} for the 
`classical' dSphs. Investigation of the stellar contamination issue
is done by comparing reconstructed $J$-factors to their true values,
for a set of mock data presenting different levels of Milky Way or
stream contamination. Doing so, we find that:

\begin{itemize}
   \item $J$-factors can be robustly reconstructed when large enough
     samples of stars are available (`classical' dSphs), whereas small
     samples (`ultrafaint' dSphs) are more sensitive to contamination,
     with their $J$- and $D$-factors more likely to overshoot the true
     value;
   \item discrepant $J$- or $D$-factors from the $P_i$-weighted and
     $P_i^{>0.95}$-cut analyses hint at high levels of
     contamination. The $P_i^{>0.95}$-cut analysis is found to give
     generally more conservative results (large CIs, but encompassing
     the true value) while the $P_i$-weighted analysis tends to
     overshoot (small CIs, true value outside CIs).
\end{itemize}

Therefore, in the remainder of the paper our results are based only on
stars with $P_i > 0.95$, whenever this information is available. A
direct consequence of this is that Segue~I (among the most favoured targets)
may become one of the least reliable targets to set constraints on DM (Bonnivard, Maurin \& Walker in prep.).
Unsurprisingly this confirms that the `classical' dSphs are
the most robust targets as they are little affected by contamination.

\section{Results}
\label{sec:Results}
Using the MCMC analysis described in the previous Section, we fit the
velocity data of the 8 `classical' and of 13 `ultrafaint' dSph
galaxies with the 7 parameters (3 parameters for the Einasto DM
profile and 4 for the Baes \& Van Hese velocity anisotropy) required
in our Jeans analysis.  For each point in the chains, any relevant
quantity may be computed and its median value and credibility
intervals estimated from the resulting distribution.

This is true of the reconstructed velocity dispersion profile of each
dSph, the median and 95\% CIs of which are plotted in solid and dashed
blue lines in figures~\ref{fig:disp_classical} and
\ref{fig:disp_ultrafaint}.\footnote{For display purposes, the binned
  analysis has been used in these figures, i.e. binned data and $\cal
  L^{\rm bin}$ likelihood function. We have checked for each dSph that the results obtained using the binned or unbinned analysis are consistant.} The reconstructed profiles and
CIs appear to always provide a good representation of the data, with
much wider CIs for `ultrafaint' dSphs (up to a factor $\sim 5$ at
large radii, compared to a factor $\sim 2$ for `classical' dSphs)
because of the sparsity of their stellar data.

\begin{figure}
\includegraphics[width=\columnwidth]{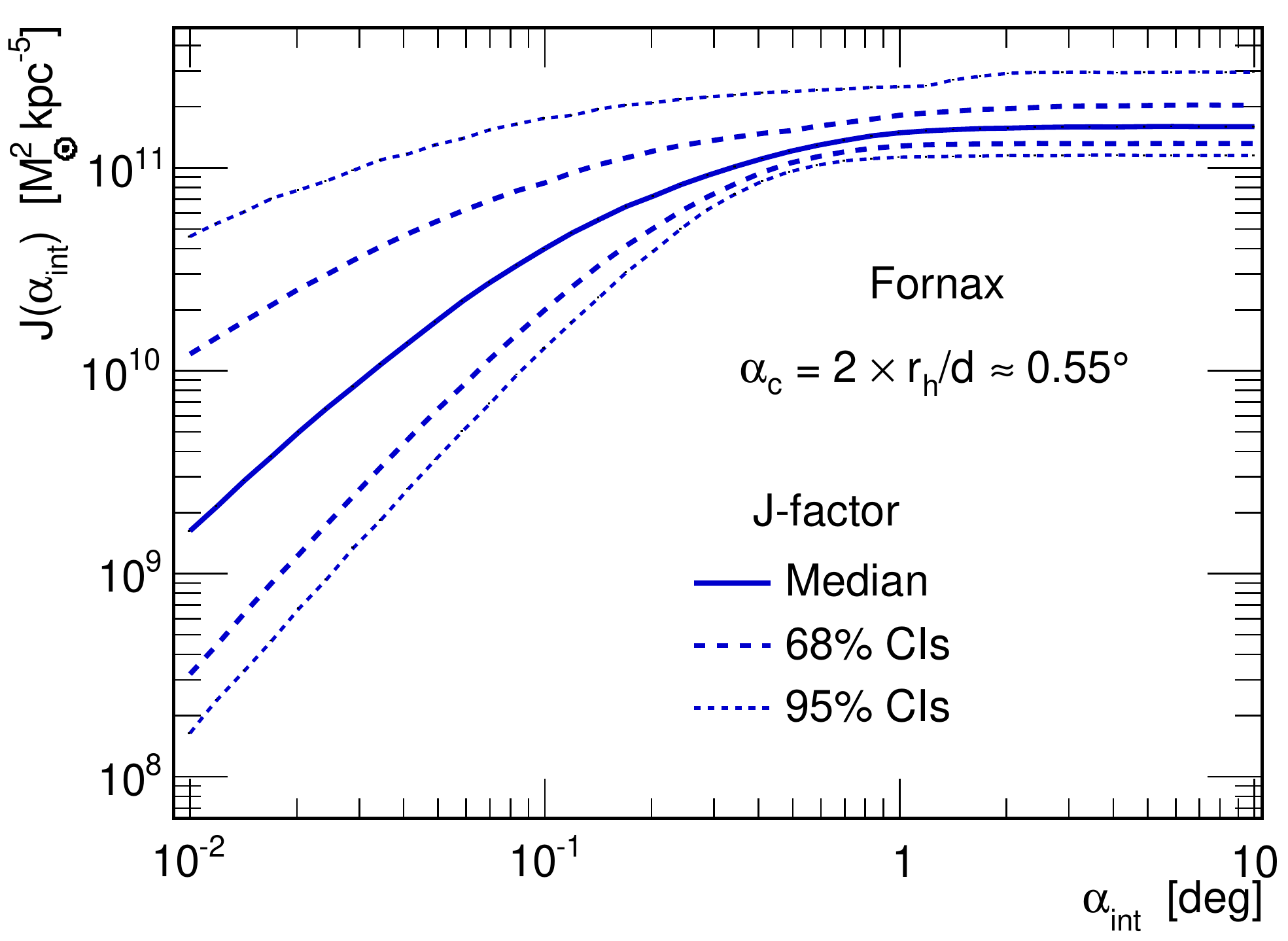}
\includegraphics[width=\columnwidth]{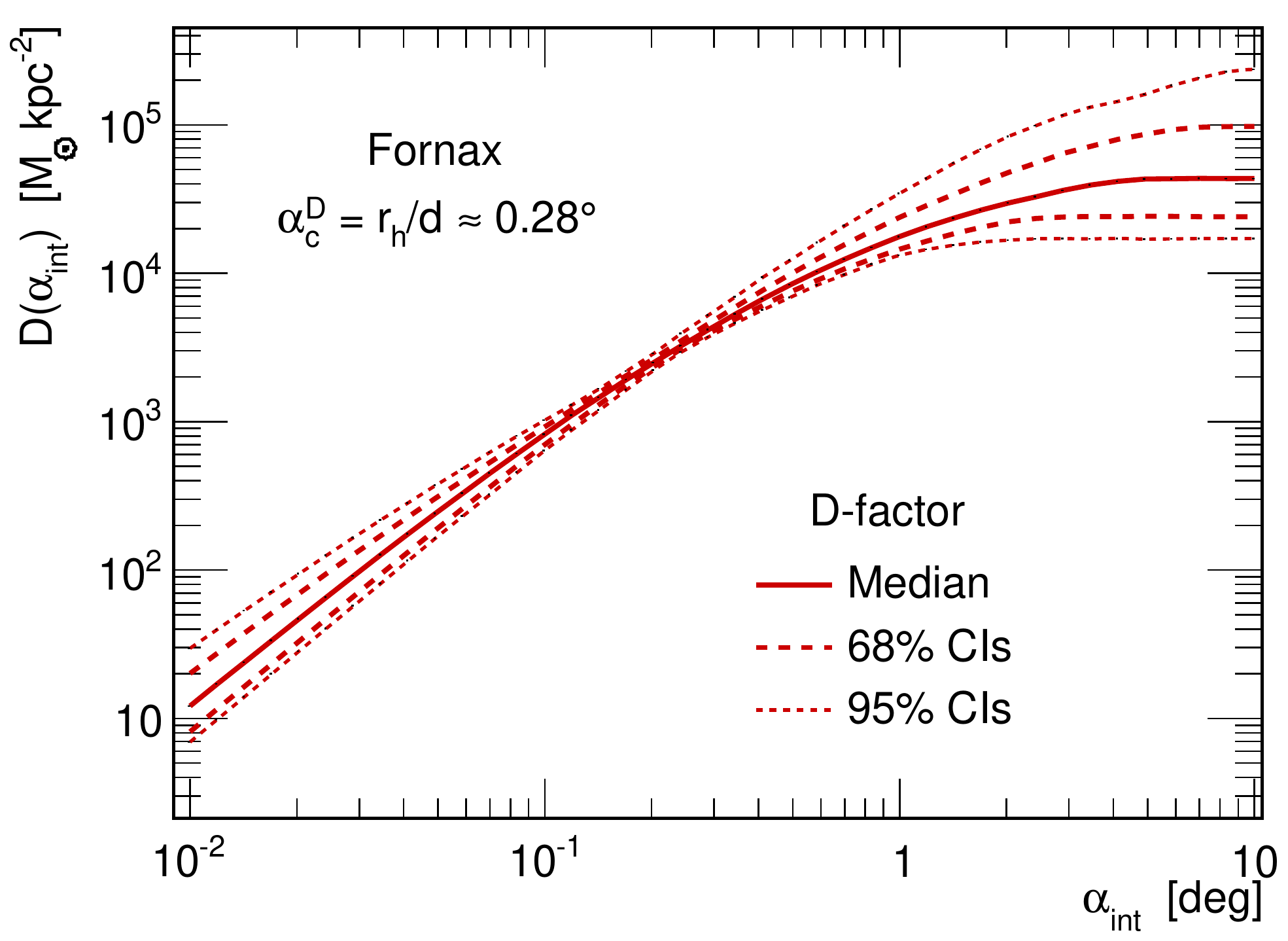}
\caption{$J$- and $D$-factor (top and bottom panel respectively) as a
  function of the integration angle $\alpha_{\text{int}}$ for the
  `classical' dSph Fornax.}
\label{fig:J_Fornax}
\end{figure}

The three Einasto profile parameters are used to compute the DM
annihilation $J$-factors and decaying DM $D$-factors using
equation~(\ref{eq:J}). Figure~\ref{fig:J_Fornax} displays the median
value and 95\% CIs of the $J$- and $D$-factors, for the `classical'
dSph Fornax, as a function of $\alpha_{\rm int}$.  Once the
integration angle encompasses the whole halo, the astrophysical
factors saturate. This figure shows the optimal integration angle
$\alpha_c$ for which the CIs for the $J$- and $D$-factors are the
smallest, i.e., where these astrophysical factors may be robustly
determined despite our inability to constrain the inner slope of their
DM profile. For Fornax, $\alpha_c^D\sim 0.28^\circ$ as may be seen
from the figure. {\tt ASCII} files of the median values, 68\% and 95\%
CIs of $J(\alpha_{\rm int})$ and $D(\alpha_{\rm int})$ for all the
dSphs discussed in this paper may be retrieved from the Supporting
Information submitted with this paper.
\subsection{J and D-factors of dSphs vs Galactic background}
\label{subsec:gal_bckg}
All the dSph satellite galaxies of the MW are embedded in its DM
halo. Hence in both the annihilating and decaying DM scenario, our
Galaxy's DM halo will provide a background signal of the same nature
as that of the targeted dSph galaxy. This consideration is quite
independent of any diffuse $\gamma$-ray emission of astrophysical
origin which we do not discuss here. For simplicity, we also ignore the
extragalactic signal originating from DM annihilations or decays on cosmological
scales (and integrated over all redshifts). 

The MW halo's astrophysical $J$- and $D$-factors are computed with the
{\sc Clumpy} code, assuming the following characteristics:
\begin{itemize}
\item We use an Einasto profile to model the smooth DM distribution,
  scaled to the local DM density ($\rho_\odot=0.3$\;GeV\;cm$^{-3}$);
\item We include the contribution of a population of DM clumps, having
  a cored spatial distribution and a mass distribution $dN/dM\propto
  M^{-1.9}$ \citepads{2008MNRAS.391.1685S} (amounting to $\sim10\%$ of
  the MW's total mass);
\item We assume that the mass-concentration relation\footnote{The
  mass-concentration relation is a fundamental ingredient to relate a
  halo mass and size to the scale density and radius required by the
  DM density parametrisation.} has a log-normal distribution
  \citepads{2014MNRAS.442.2271S}.
\end{itemize}

\begin{figure*}
\includegraphics[width=\columnwidth]{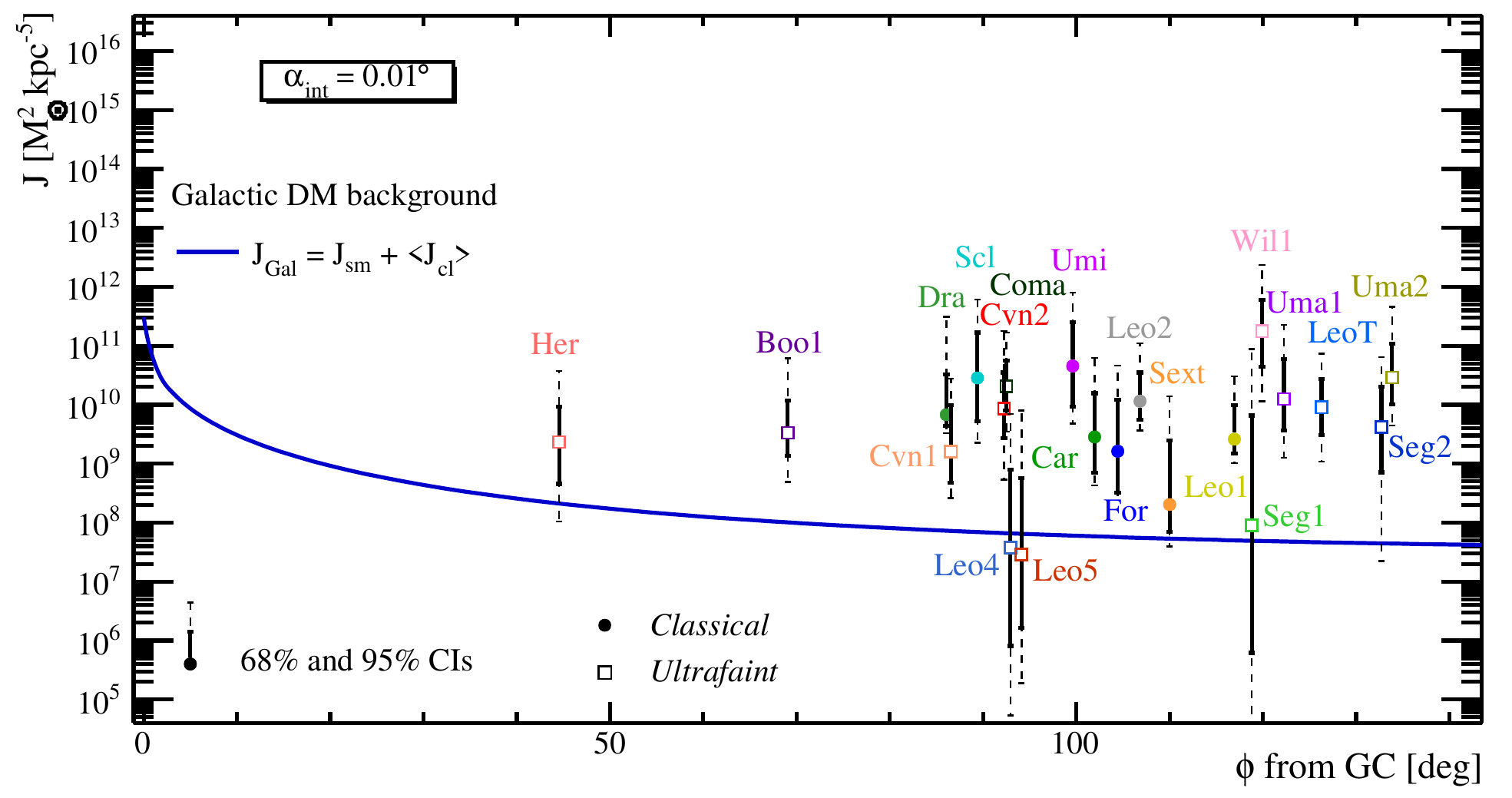}
\includegraphics[width=\columnwidth]{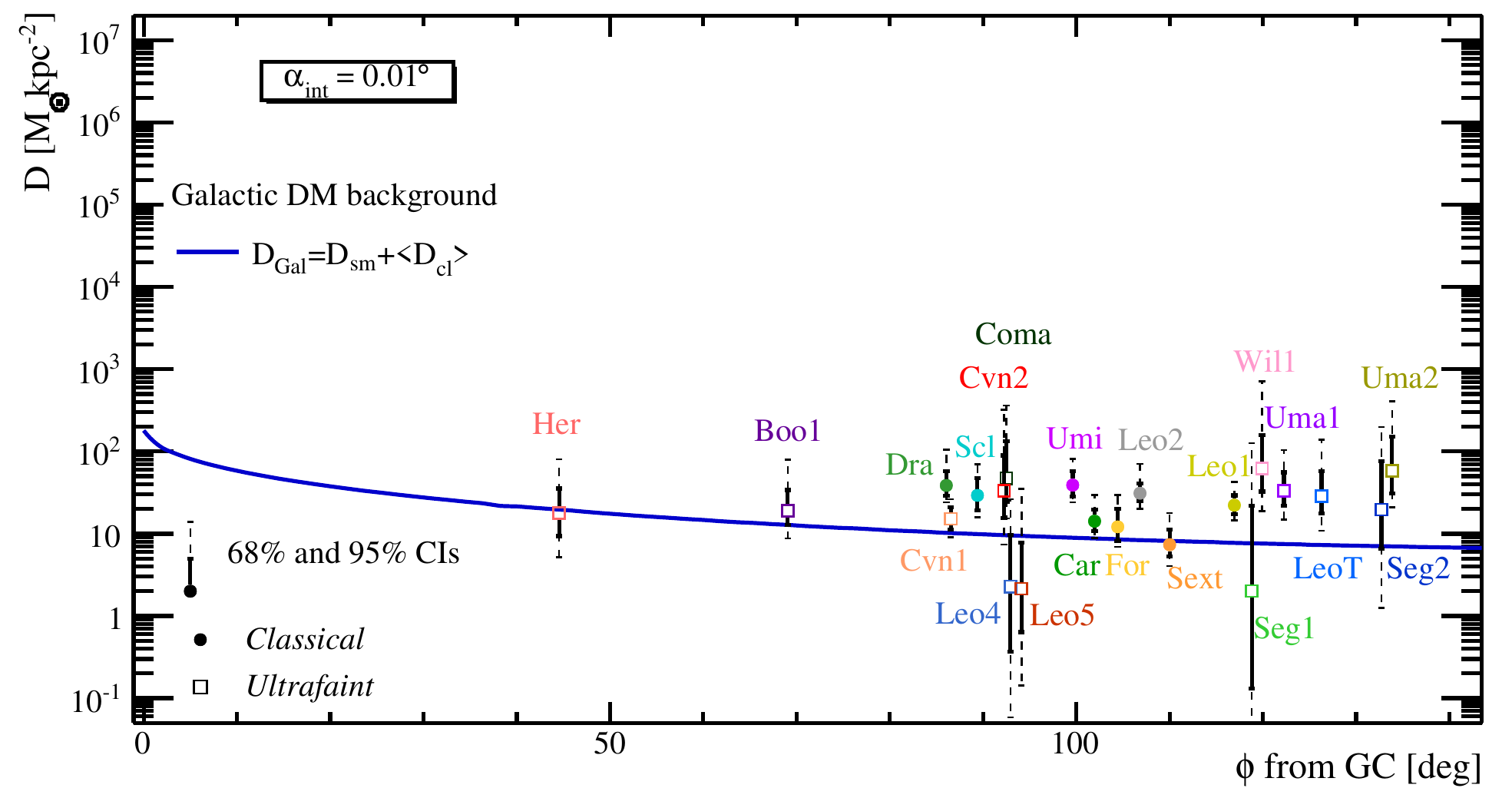}
\includegraphics[width=\columnwidth]{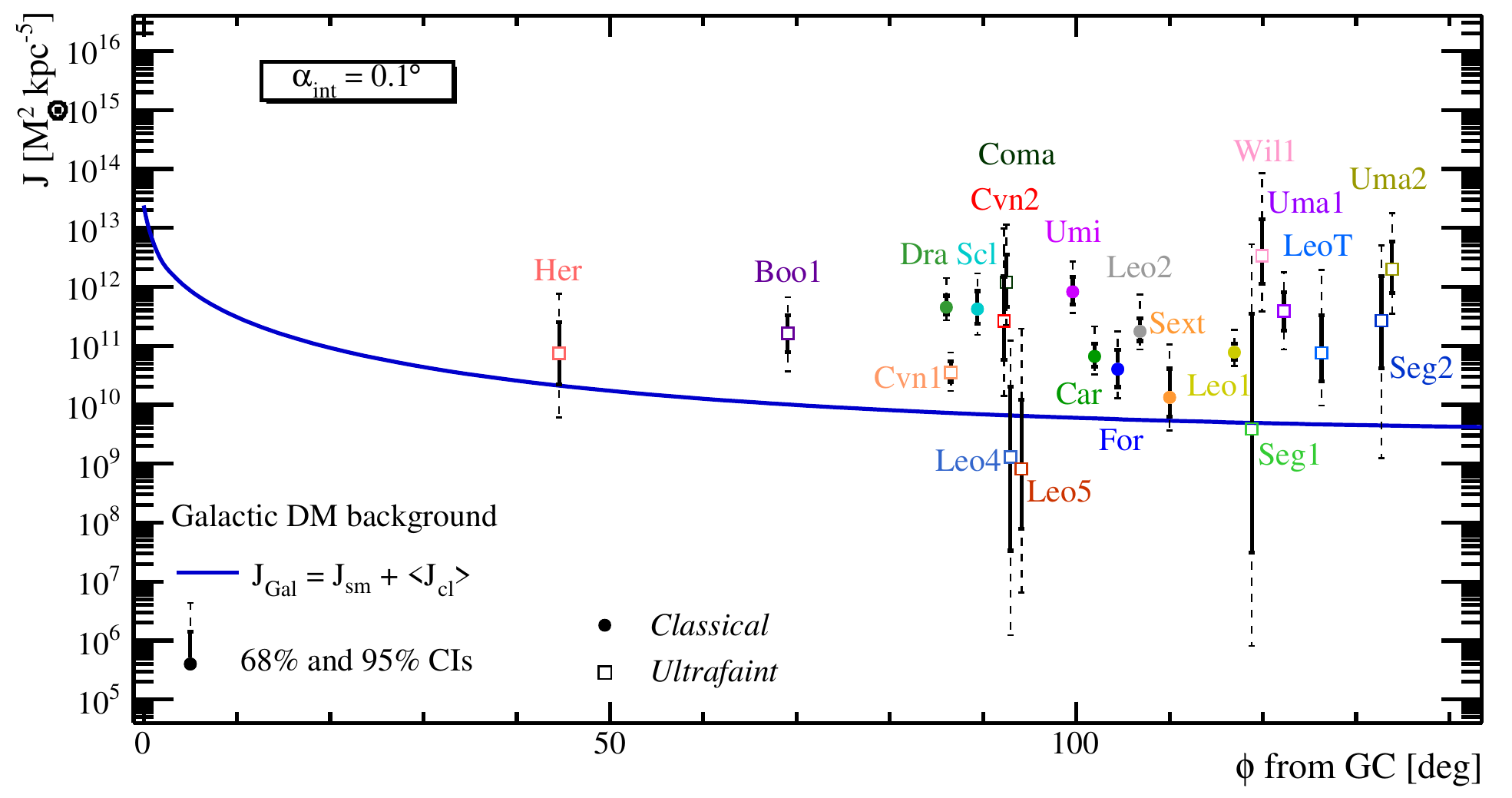}
\includegraphics[width=\columnwidth]{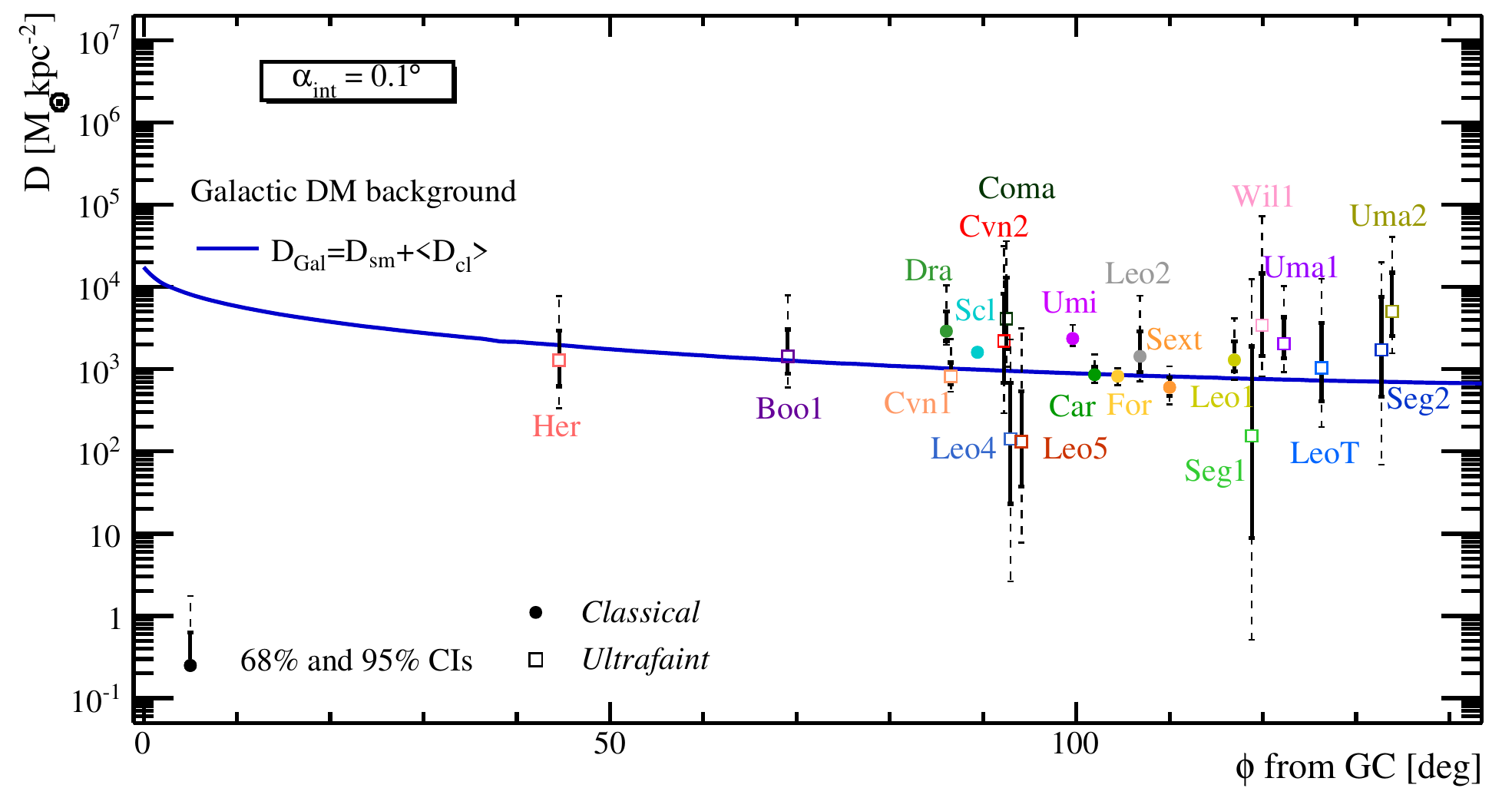}
\includegraphics[width=\columnwidth]{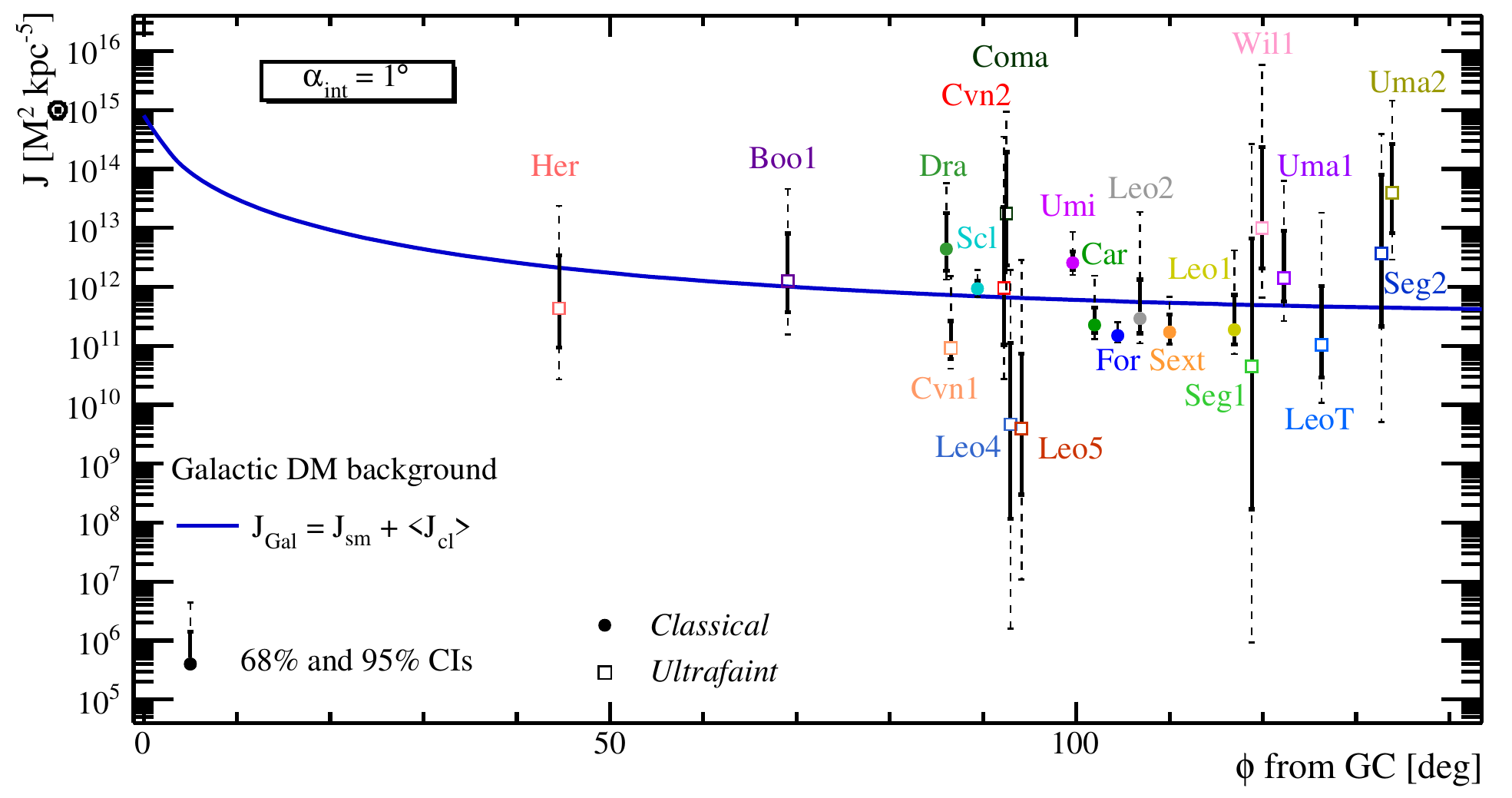}
\includegraphics[width=\columnwidth]{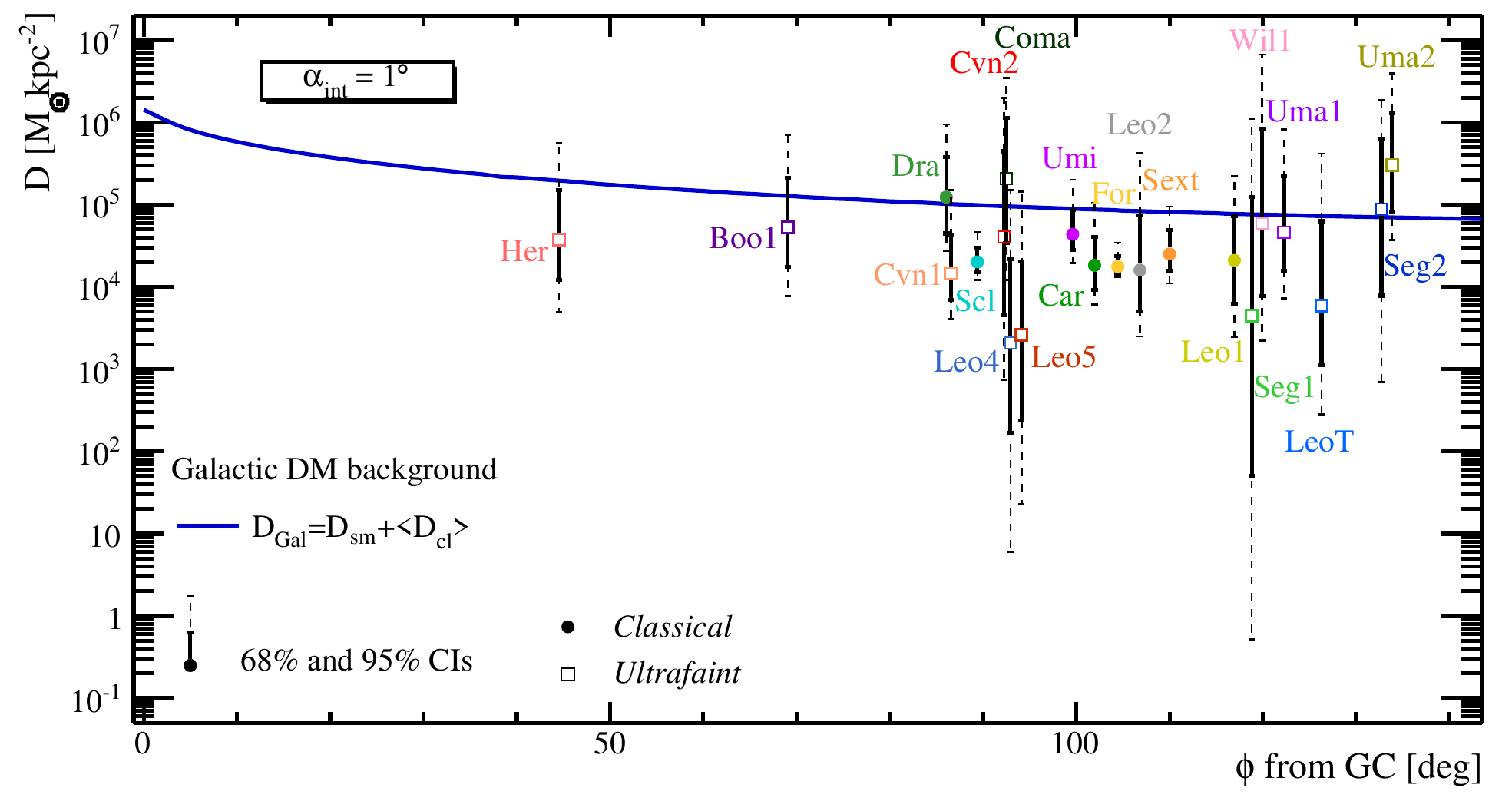}
\caption{Annihilation (left) and decay (right) factors of the 21 dSphs
  studied (symbols) versus the Galactic DM halo background (blue
  line), as function of the angular distance from the Galactic
  centre. Top to bottom panels correspond to three integration angles:
  $0.01^{\circ}$, $0.1^{\circ}$, and $1^{\circ}$. See
  Section~\ref{subsec:gal_bckg} for details.}
\label{fig:J}
\end{figure*}
Figure~\ref{fig:J} displays the $J$- (left) and $D$-factors (right) of
the 21 dSph galaxies (symbols) studied in this paper as a function of
their angular distance from the Galactic centre. The solid blue line
corresponds to our estimate of the contribution from the MW DM
halo. This has been repeated for three integration angles,
$\alpha_{\rm int}=0.01^\circ, 0.1^\circ, 1^\circ$. This figure clearly
illustrates the loss of contrast between the dSph target and the MW
background as the integration angle is increased. Indeed, the
background (exotic or not) is $\propto \alpha_{\rm int}^2$; this is
not so for the dSphs where the astrophysical factor is very centrally
peaked, especially for $J$. For the $J$-factor, most of the dSphs
appear an order of magnitude or more above the background for
$\alpha_{\rm int} = 0.01^\circ$, while this is true only of a couple
of them for $\alpha_{\rm int} = 1^\circ$. For decay (right column),
even for small integration angles, the contrast is always smaller than that
of annihilation. Therefore, for large integration angles which may
be dictated by instrumental resolution, it would be a better strategy to look
directly for the MW's DM halo signature, rather than at a specific
dSph galaxy.

\subsection{J-factor: ranking of the dSphs and comparison to other works}
\label{subsec:J-ranking}
\begin{figure*}
\includegraphics[width=1.\linewidth]{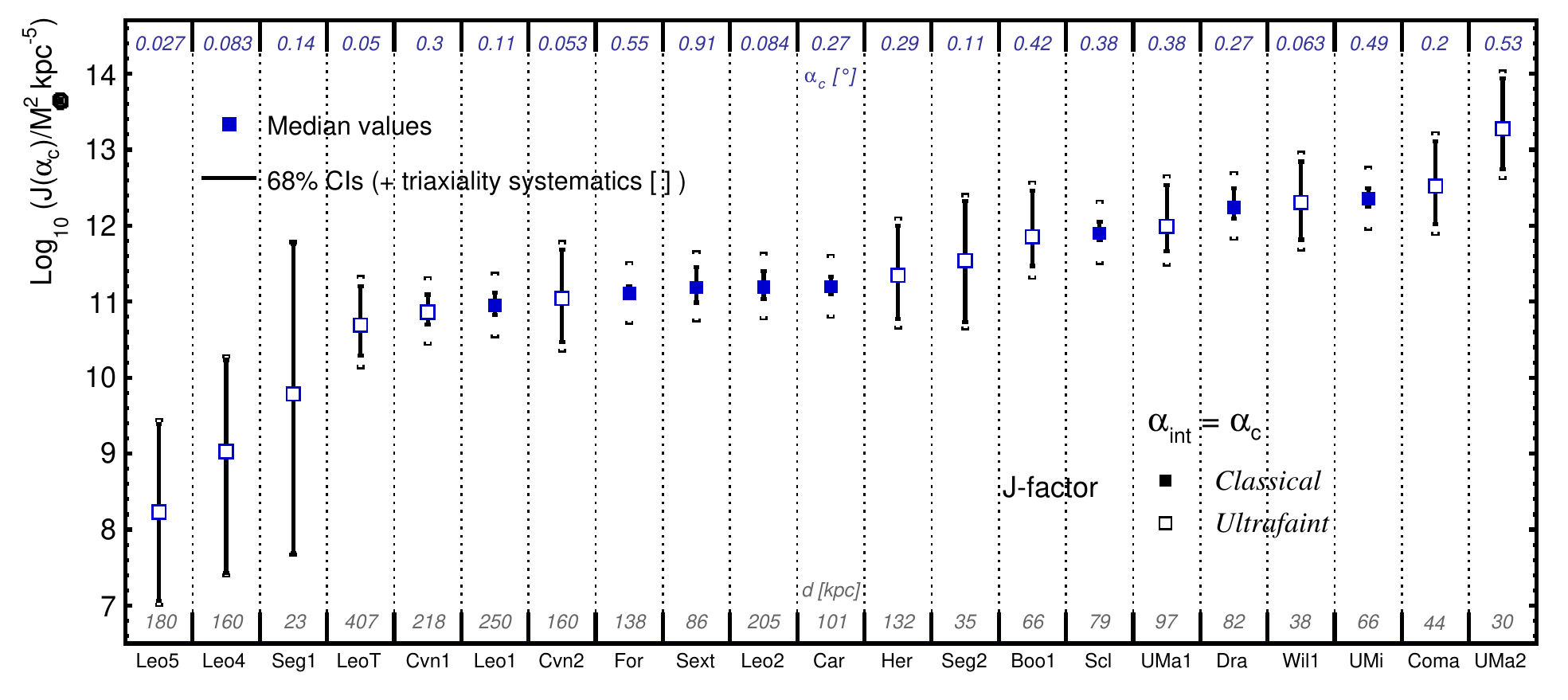}
\includegraphics[width=1.\linewidth]{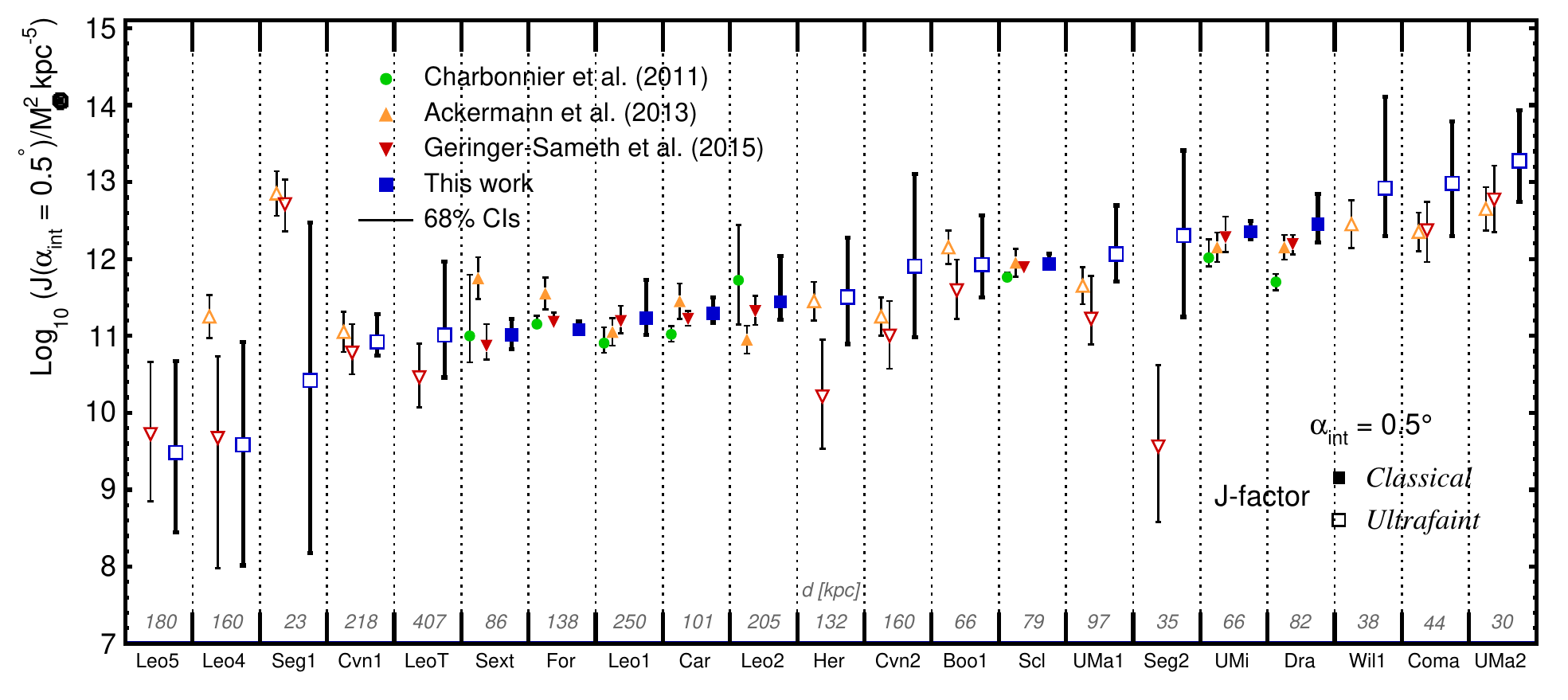}
\caption{\emph{Top}: $J$-factors and 68\% CIs for $\alpha_{\text{int}}
  = \alpha_{c}^{J}$: the '[]' symbols combine in quadrature the 68\% statistical
  uncertainties and possible systematics ($\pm0.4$) from triaxiality
  of the dSph galaxies \citepads{2015MNRAS.446.3002B}.  \emph{Bottom}:
  comparison of the J-factors to other works, with $\alpha_{\rm
    int}=0.5^\circ$. See also Section~\ref{subsec:target_by_target}
    for a critical discussion of the targets most favoured by our analysis.}
\label{fig:J_results}
\end{figure*}

Putting aside the notion of contrast mentioned above, the
astrophysical factors are the relevant proxies to determine whether or
not a given dSph is potentially interesting for indirect detection.
Table~\ref{tab:results} summarises our findings and we compare these
results in figure~\ref{fig:J_results} to other studies in the
literature.

\paragraph*{Ranking at $\alpha_{\rm int}=\alpha_c\approx2r_h /d$.} \rm
The top panel in figure~\ref{fig:J_results} shows the $J$-factors
ordered according to their median from the faintest (Leo5) to the
brightest (UMa2), when integrating the signal up to the optimal angle
$\alpha_{\rm int}=\alpha_c$, where the $J$-factors have the smallest
error bars. To account for possible systematics from triaxiality of the DM halo
(which depends on the l.o.s. orientation of each dSph, see \citealtads{2015MNRAS.446.3002B}), the error bars ('[]' symbols) combine a 0.4 dex
uncertainty in quadrature with the 68\% CIs. Our updated analysis (compared to
the results presented in \citealp{2011MNRAS.418.1526C}) still prefers
UMi among all the `classical' dSph while Coma and Ursa Major 2 are the
most promising `ultrafaint' dSphs (Segue~I falls among the least
interesting targets). For each dSph galaxy, the optimal angle
$\alpha^J_c$ is quoted above the data point, while the distance to the
dSph is quoted below. Unsurprisingly, the most promising galaxies are
also among the closest. The estimated uncertainties on this plot
provide a clear signature of our data-driven approach: the
`ultrafaint' dSphs have significantly larger error bars than their
`classical' counterparts.

\paragraph*{Ranking at $\alpha_{\rm int}=0.5^\circ$ 
and comparison to other works.} \rm The bottom panel in
figure~\ref{fig:J_results} compares our results to other existing
studies for a fixed integration angle $\alpha_{\rm
  int}=0.5^\circ$, typical of the Fermi-LAT angular resolution in the GeV range. 
  First, comparing the top and bottom panels shows no
drastic ordering changes among the dSph galaxies but for a few
inversions. Second, the overall trend appears to be preserved between
the different studies, with the same objects having the highest $J$
values. Nonetheless, a closer inspection shows significant differences
both in the values of the $J$-factors and the size of the error
bars. The differences observed between our analysis and others is
understood as follows:
\begin{itemize}

\item In \citetads{2011MNRAS.418.1526C}, we conducted the same
  exercise on the 8 `classical' dSph galaxies (green circles), but
  with a more constrained Jeans analysis (using a constant anisotropy
  profile and a Plummer light profile). UMi was found to be the most
  promising target among the `classical' dSph, while Leo~2 had the
  highest median $J(\alpha_c)$ but with much larger uncertainties.
  For all the objects but Leo~2 (which is very uncertain anyway), the
  \citetads{2011MNRAS.418.1526C} values are slightly lower than their
  updated version. Changing the anisotropy and light profiles to more
  flexible parametrisations in the current analysis is the main reason
  for the differences between this and our previous study.  Overall,
  we find larger $J$-factors whenever the light data require an outer
  slope steeper than that given by the Plummer profile used in
  \citetads{2011MNRAS.418.1526C}. Note that this effect was already
  observed on mock data in \citetads{2015MNRAS.446.3002B}. The most
  striking example is the case of Draco.

\item The Fermi collaboration \citepads{2014PhRvD..89d2001A,2015arXiv150302641F} reported
  the $\gamma$-ray observations of 25 dSph galaxies and conducted a
  stacking analysis of 15 of them to set constraints on the DM
  annihilation cross section. These authors provide $J$-factors of 18
  dSph galaxies overlapping with our sample (orange triangles),
  obtained using a two-level Bayesian hierarchical modelling that
  constraints the entire population of MW dSphs simultaneously
  \citepads{2013arXiv1309.2641M}.  Their values do not show any
  particular trend when compared to ours, with most CIs
  overlapping. The most striking difference concerns the size of their
  error bars, which remain roughly the same regardless of the nature
  of the object (`classical' or `ultrafaint'). This is very likely
  related to their two-level hierarchical analysis, where the entire
  population of MW dSph galaxies is to some extent assumed to share
  the same properties.  The constraints coming from the `classical'
  dSphs are therefore redistributed to the `ultrafaint' dSphs. Note
  also that a NFW profile for the DM density was assumed, but the
  analysis found to be fairly insensitive to this choice.

\item In \citetads{2015ApJ...801...74G}, the authors provide $J$- and
  $D$-factors of 20 dSph galaxies obtained using a data-driven Jeans
  analysis quite similar to ours.  In particular, they use the
  same unbinned likelihood as in
  equation~(\ref{eq:likelihood_unbinned}), but with a Zhao DM density
  profile, a constant anisotropy and a Plummer light profile.
  However, they select the radius of the outermost observed star as
  truncation radius for computing the astrophysical factors, which
  results in lower values (red triangles) than ours (blue
  squares). Segue~II (Seg2), Hercules (Her), and Ursa Major~I
  (UMa1) are particularly affected by this choice, with
  $R_\text{max}/r_\text{h}$ values between $\sim 2$ and $\sim 4$,
  while we obtain $R_\text{max}/r_\text{h}$ values $\gtrsim 10$ with
  our estimations of the halo size (see Section
  \ref{subsec:size}). Note that the underestimation of the halo size
  can also lead to an underestimation of the CIs (Appendix
  \ref{app:rvir}).  This partially explains the larger error bars we
  find, which are also a consequence of the more flexible anisotropy
  and light profiles parametrisations we use
  \citepads{2015MNRAS.446.3002B}.
\end{itemize}

In summary, figure~\ref{fig:J_results} (bottom) highlights the fact
that the values found for $J$-factors and their CIs can depend
strongly on the underlying assumptions. We believe our present work to
better reflect the various sources of uncertainties that affect
$J$-factor estimations, following the thorough validation of the
method initiated in \citetads{2015MNRAS.446.3002B} and concluded in
Appendices~\ref{app:unbinned} and \ref{app:rvir}.  Our data-driven
analysis naturally implies larger error bars for dSphs with fewer
stars, and vice-versa. Also, thanks to a more flexible parametrisation
of light profile, we find higher $J$-factors for some targets, which
may lower further the upper limits for the DM annihilation cross
section
\citepads{2014PhRvD..89d2001A,2014arXiv1410.2242G,2015arXiv150302641F}.

\begin{table*}
  \caption{Summary of results for the 21 dSph galaxies presented in this
    paper. The dSphs are ordered by distance and the
    columns correspond to (from left to right): name, distance,
    optimal angle for annihilation ($\alpha_c^J \approx2
    \alpha_c^D\approx2r_h /d$), median $J$-factors and 68\% (95\%) CIs
    for $\alpha_{\rm int}=0.01,\,0.5,\,\alpha_c^J$, median $D$-factors
    and 68\% (95\%) CIs for $\alpha_{\rm
      int}=0.01,\,0.5,\,\alpha_c^D$. Note that a systematic
    uncertainty of $\pm0.4$ (resp.~$\pm0.3$) must be allowed 
    in order to the reflect the possible triviality of the dSph galaxies
    \citepads{2015MNRAS.446.3002B}.}
\label{tab:results}
\begin{tabular}{@{}lccccccccc} \hline\hline
  dSph    &   d    & $\alpha_c^J$  &$\log_{10}[J(0.01^\circ)]$   &  $\log_{10}[J(0.5^\circ)]$     &     $\log_{10}[J(\alpha_c^J)]$ \!\!&& \!\!$\log_{10}[D(0.01^\circ)]$\!\!   &  $\log_{10}[D(0.1^\circ)]$     &     \!\!$\log_{10}[D(\alpha_c^D)]$\!\!\\
          & \!\!\![kpc]\!\!\!  &\!\!\![$\deg$]\!\!\! &\multicolumn{3}{c}{$[J/M_{\odot}^{2}\,$kpc$^{-5}]$} &&\multicolumn{3}{c}{$[D/M_{\odot}\,$kpc$^{-2}]$}\vspace{0.05cm}\\ 
\hline 
{\it Segue\;I}                      &  23 & 0.14 &$~8.0_{-2.2(-3.5)}^{+1.9(+3.0)}$ & $10.4_{-2.2(-4.5)}^{+2.1(+3.5)}$ &  $~9.8_{-2.1(-3.8)}^{+2.0(+3.2)}$ && $0.3_{-1.2(-2.1)}^{+1.0(+1.8)}$ & $2.2_{-1.2(-2.5)}^{+1.1(+1.9)}$ &  $1.9_{-1.2(-2.3)}^{+1.1(+1.9)}$ \vspace{0.2cm}\\
{\it Ursa Major\;II}\!\!\!\!\!\!    &  30 & 0.53 &$10.5_{-0.5(-0.8)}^{+0.6(+1.2)}$ & $13.3_{-0.5(-0.9)}^{+0.7(+1.3)}$ &  $13.4_{-0.6(-1.0)}^{+0.7(+1.4)}$ && $1.8_{-0.3(-0.4)}^{+0.4(+0.8)}$ & $3.7_{-0.3(-0.5)}^{+0.5(+0.9)}$ &  $4.6_{-0.4(-0.6)}^{+0.5(+1.0)}$ \vspace{0.2cm}\\
{\it Segue\;II}                     &  35 & 0.11 &$~9.6_{-0.8(-2.3)}^{+0.7(+1.2)}$ & $12.3_{-1.1(-2.7)}^{+1.1(+1.7)}$ &  $11.5_{-0.8(-2.3)}^{+0.8(+1.3)}$ && $1.3_{-0.5(-1.2)}^{+0.6(+1.0)}$ & $3.2_{-0.6(-1.4)}^{+0.6(+1.1)}$ &  $2.8_{-0.5(-1.3)}^{+0.6(+1.0)}$ \vspace{0.2cm}\\
{\it Willman\;I}                     &  38 & 0.06 &$11.2_{-0.6(-1.2)}^{+0.5(+1.1)}$ & $12.9_{-0.6(-1.1)}^{+1.2(+2.3)}$ &  $12.3_{-0.5(-1.0)}^{+0.5(+1.2)}$ && $1.8_{-0.3(-0.5)}^{+0.4(+1.1)}$ & $3.5_{-0.4(-0.6)}^{+0.6(+1.3)}$ &  $2.8_{-0.3(-0.5)}^{+0.5(+1.2)}$ \vspace{0.2cm}\\
{\it Coma}                          &  44 & 0.20 &$10.3_{-0.4(-0.8)}^{+0.4(+0.9)}$ & $13.0_{-0.7(-1.1)}^{+0.8(+1.4)}$ &  $12.5_{-0.5(-0.8)}^{+0.6(+1.1)}$ && $1.7_{-0.3(-0.5)}^{+0.4(+0.9)}$ & $3.6_{-0.4(-0.6)}^{+0.5(+0.9)}$ &  $3.6_{-0.4(-0.6)}^{+0.5(+0.9)}$ \vspace{0.2cm}\\
Ursa Minor\!\!\!\!\!\!              &  66 & 0.49 &$10.7_{-0.7(-1.0)}^{+0.7(+1.2)}$ & $12.4_{-0.1(-0.2)}^{+0.1(+0.4)}$ &  $12.4_{-0.1(-0.2)}^{+0.1(+0.4)}$ && $1.6_{-0.1(-0.2)}^{+0.2(+0.3)}$ & $3.3_{-0.0(-0.1)}^{+0.1(+0.2)}$ &  $4.0_{-0.1(-0.2)}^{+0.1(+0.3)}$ \vspace{0.2cm}\\
{\it Bo\"otes\;I}                   &  66 & 0.42 &$~9.5_{-0.4(-0.8)}^{+0.5(+1.3)}$ & $11.9_{-0.4(-0.8)}^{+0.6(+1.2)}$ &  $11.9_{-0.4(-0.7)}^{+0.6(+1.1)}$ && $1.3_{-0.2(-0.3)}^{+0.3(+0.6)}$ & $3.2_{-0.2(-0.4)}^{+0.3(+0.7)}$ &  $3.7_{-0.3(-0.5)}^{+0.4(+0.8)}$ \vspace{0.2cm}\\
Sculptor                            &  79 & 0.38 &$10.4_{-0.7(-1.1)}^{+0.8(+1.3)}$ & $11.9_{-0.1(-0.1)}^{+0.1(+0.3)}$ &  $11.9_{-0.1(-0.1)}^{+0.1(+0.4)}$ && $1.5_{-0.2(-0.3)}^{+0.2(+0.4)}$ & $3.2_{-0.0(-0.1)}^{+0.0(+0.1)}$ &  $3.6_{-0.0(-0.1)}^{+0.0(+0.1)}$ \vspace{0.2cm}\\
Draco                               &  82 & 0.28 &$~9.8_{-0.2(-0.3)}^{+0.7(+1.7)}$ & $12.5_{-0.2(-0.4)}^{+0.4(+0.7)}$ &  $12.2_{-0.1(-0.3)}^{+0.3(+0.6)}$ && $1.6_{-0.1(-0.2)}^{+0.2(+0.4)}$ & $3.5_{-0.1(-0.2)}^{+0.2(+0.5)}$ &  $3.8_{-0.2(-0.3)}^{+0.3(+0.6)}$ \vspace{0.2cm}\\
Sextans                             &  86 & 0.91 &$~8.3_{-0.5(-0.7)}^{+1.1(+1.8)}$ & $11.0_{-0.2(-0.4)}^{+0.2(+0.4)}$ &  $11.2_{-0.2(-0.4)}^{+0.3(+0.6)}$ && $0.9_{-0.1(-0.3)}^{+0.2(+0.4)}$ & $2.8_{-0.1(-0.2)}^{+0.1(+0.3)}$ &  $4.0_{-0.1(-0.3)}^{+0.2(+0.4)}$ \vspace{0.2cm}\\
{\it Ursa Major\;I}\!\!\!\!\!\!     &  97 & 0.38 &$10.1_{-0.5(-1.0)}^{+0.7(+1.3)}$ & $12.1_{-0.4(-0.7)}^{+0.6(+1.3)}$ &  $12.0_{-0.3(-0.6)}^{+0.6(+1.2)}$ && $1.5_{-0.2(-0.4)}^{+0.2(+0.5)}$ & $3.3_{-0.2(-0.3)}^{+0.3(+0.7)}$ &  $3.8_{-0.2(-0.4)}^{+0.4(+0.8)}$ \vspace{0.2cm}\\
Carina                              & 101 & 0.27 &$~9.4_{-0.6(-0.8)}^{+0.7(+1.3)}$ & $11.3_{-0.1(-0.2)}^{+0.2(+0.5)}$ &  $11.2_{-0.1(-0.2)}^{+0.1(+0.4)}$ && $1.2_{-0.1(-0.2)}^{+0.1(+0.3)}$ & $2.9_{-0.1(-0.1)}^{+0.1(+0.2)}$ &  $3.2_{-0.1(-0.1)}^{+0.1(+0.3)}$ \vspace{0.2cm}\\
{\it Hercules}               & 132 & 0.29 &$~8.6_{-0.6(-1.3)}^{+0.6(+1.3)}$ & $10.9_{-0.7(-1.2)}^{+0.7(+1.6)}$ &  $10.7_{-0.6(-1.1)}^{+0.7(+1.4)}$ && $0.9_{-0.3(-0.6)}^{+0.3(+0.7)}$ & $2.8_{-0.3(-0.6)}^{+0.4(+0.8)}$ &  $3.0_{-0.3(-0.6)}^{+0.4(+0.9)}$ \vspace{0.2cm}\\
Fornax                              & 138 & 0.56 &$~9.2_{-0.7(-1.0)}^{+0.9(+1.5)}$ & $11.1_{-0.1(-0.1)}^{+0.1(+0.3)}$ &  $11.1_{-0.1(-0.1)}^{+0.1(+0.3)}$ && $1.1_{-0.2(-0.2)}^{+0.2(+0.4)}$ & $2.9_{-0.1(-0.1)}^{+0.0(+0.1)}$ &  $3.6_{-0.0(-0.0)}^{+0.0(+0.1)}$ \vspace{0.2cm}\\
{\it Leo\;IV}                       & 160 & 0.08 &$~7.6_{-1.7(-2.8)}^{+1.3(+2.3)}$ & $~9.6_{-1.6(-3.4)}^{+1.3(+2.4)}$ &  $~9.0_{-1.6(-3.0)}^{+1.2(+2.0)}$ && $0.4_{-0.8(-1.6)}^{+0.6(+1.1)}$ & $2.2_{-0.8(-1.7)}^{+0.7(+1.2)}$ &  $1.5_{-0.8(-1.6)}^{+0.6(+1.1)}$ \vspace{0.2cm}\\
{\it Canis Venatici\;II}\!\!\!\!\!\!\!& 160 & 0.05 &$~9.9_{-0.5(-1.2)}^{+0.6(+1.3)}$ & $11.9_{-0.9(-1.5)}^{+1.2(+2.3)}$ &  $11.0_{-0.6(-1.2)}^{+0.6(+1.3)}$ && $1.5_{-0.3(-0.7)}^{+0.4(+1.0)}$ & $3.3_{-0.5(-0.9)}^{+0.6(+1.2)}$ &  $2.3_{-0.4(-0.7)}^{+0.5(+1.0)}$ \vspace{0.2cm}\\
{\it Leo\;V}                        & 180 & 0.027&$~7.5_{-1.2(-2.2)}^{+1.3(+2.4)}$ & $~9.5_{-1.0(-2.5)}^{+1.2(+2.7)}$ &  $~8.2_{-1.2(-2.1)}^{+1.2(+2.4)}$ && $0.3_{-0.5(-1.2)}^{+0.6(+1.2)}$ & $2.1_{-0.5(-1.2)}^{+0.6(+1.4)}$ &  $0.6_{-0.5(-1.2)}^{+0.6(+1.2)}$ \vspace{0.2cm}\\
Leo\;II                             & 205 & 0.08 &$10.1_{-0.3(-0.5)}^{+0.5(+1.0)}$ & $11.4_{-0.2(-0.4)}^{+0.6(+1.5)}$ &  $11.2_{-0.2(-0.3)}^{+0.2(+0.5)}$ && $1.5_{-0.1(-0.2)}^{+0.1(+0.4)}$ & $3.2_{-0.2(-0.3)}^{+0.3(+0.7)}$ &  $2.6_{-0.1(-0.2)}^{+0.2(+0.6)}$ \vspace{0.2cm}\\
{\it Canis Venatici\;I}\!\!\!\!\!\!\! & 218 & 0.3  &$~9.2_{-0.5(-0.8)}^{+0.8(+1.2)}$ & $10.9_{-0.2(-0.3)}^{+0.4(+0.9)}$ &  $10.9_{-0.2(-0.3)}^{+0.3(+0.7)}$ && $1.2_{-0.1(-0.2)}^{+0.1(+0.2)}$ & $2.9_{-0.1(-0.2)}^{+0.2(+0.5)}$ &  $3.3_{-0.1(-0.2)}^{+0.2(+0.6)}$ \vspace{0.2cm}\\
Leo\;I                              & 250 & 0.11 &$~9.4_{-0.2(-0.4)}^{+0.6(+1.1)}$ & $11.2_{-0.2(-0.4)}^{+0.5(+1.1)}$ &  $10.9_{-0.1(-0.2)}^{+0.2(+0.4)}$ && $1.3_{-0.1(-0.2)}^{+0.1(+0.3)}$ & $3.1_{-0.1(-0.2)}^{+0.2(+0.5)}$ &  $2.7_{-0.1(-0.2)}^{+0.2(+0.4)}$ \vspace{0.2cm}\\
{\it LeoT}                          & 407 & 0.05 &$10.0_{-0.5(-0.9)}^{+0.5(+0.9)}$ & $11.0_{-0.6(-1.0)}^{+1.0(+2.1)}$ &  $10.7_{-0.4(-0.8)}^{+0.5(+1.1)}$ && $1.5_{-0.2(-0.4)}^{+0.3(+0.7)}$ & $3.0_{-0.4(-0.7)}^{+0.5(+1.1)}$ &  $2.1_{-0.3(-0.5)}^{+0.4(+0.8)}$ \\
\hline
\end{tabular}
\end{table*}

\subsection{D-factor: ranking of the dSphs and comparison to other works}
\label{subsec:D-ranking}

Dark matter decay is less often considered than annihilation, however
recent observations of an unidentified X-ray line at 3.55~keV in
galaxy clusters has generated increasing interest in this possibility
(e.g., \citealp{2014ApJ...789...13B,2014PhRvL.113y1301B}).

\begin{figure*}
\includegraphics[width=1.\linewidth]{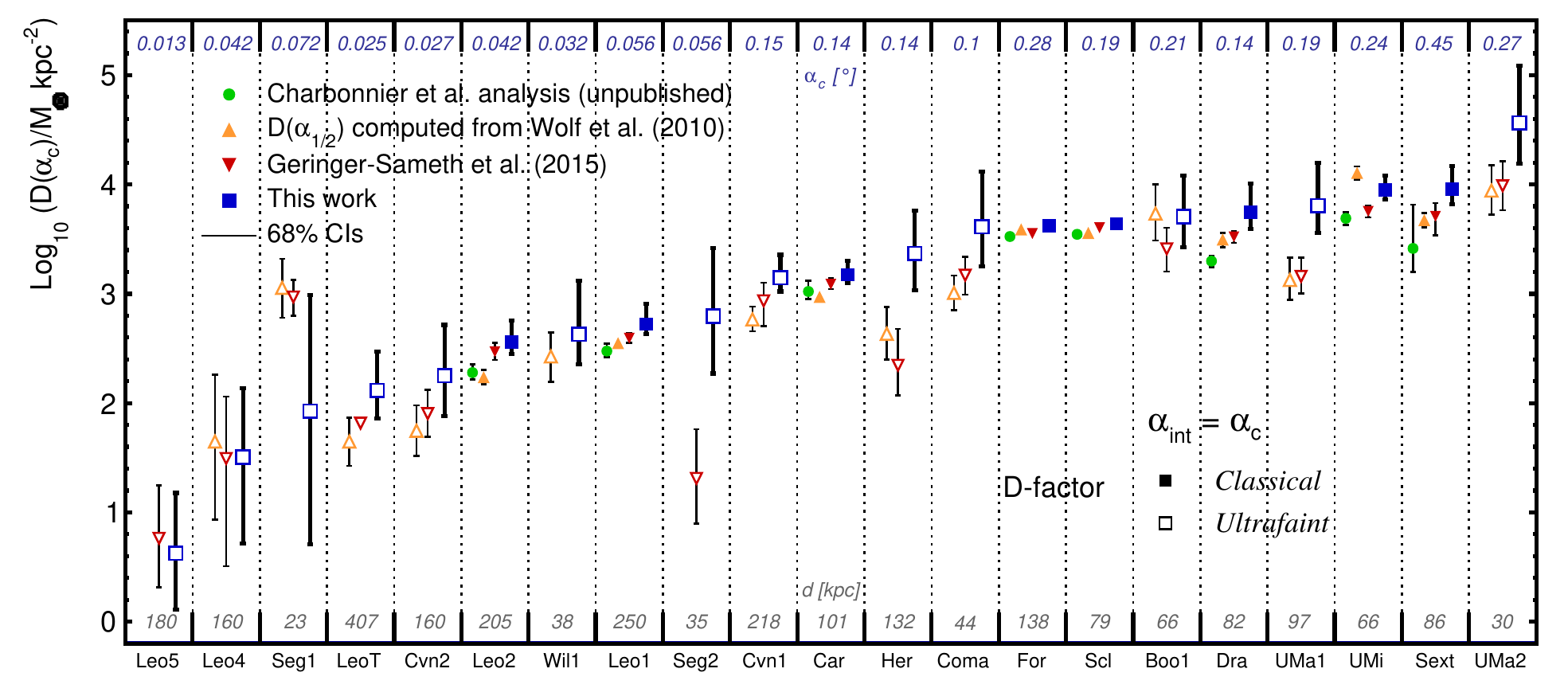}
\includegraphics[width=1.\linewidth]{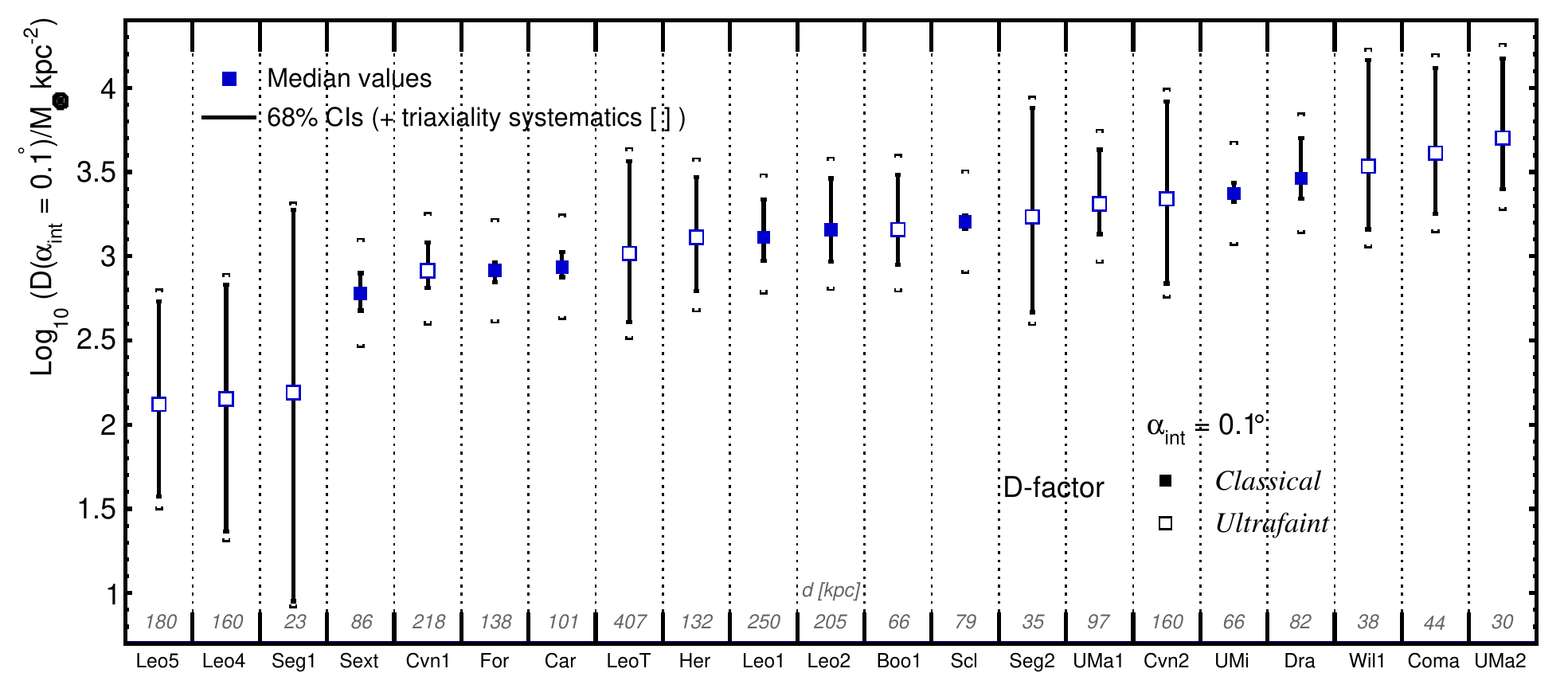}
\caption{\emph{Top}: $D$-factors and 68\% CIs for $\alpha_{\text{int}}
  = \alpha_{c}^{D}$ and comparisons to other works \emph{Bottom}: same
  but for $\alpha_{\text{int}} = 0.1^\circ$: the '[]' symbols combine
  the 68\% statistical uncertainties and possible systematics
  ($\pm0.3$) from triaxiality of the dSph galaxies
  \citepads{2015MNRAS.446.3002B}. See also Section~\ref{subsec:target_by_target}
    for a critical discussion of the targets most favoured by our analysis.}
\label{fig:D_results}
\end{figure*}
  
\paragraph*{Ranking.} The blue squares in figure~\ref{fig:D_results} and 
the three rightmost columns of table~\ref{tab:results} give an
overview of the $D$-factors computed here. First,
comparing the top panels of figures~\ref{fig:J_results}
and \ref{fig:D_results}, we find that the ordering of the most
promising targets changes significantly whether focusing on DM
annihilation or decay, even though Ursa Major\;II remains the best
candidate for $\alpha_{\rm int}=\alpha_c^D$. Furthermore, the two
panels in figure \ref{fig:D_results} show that changing the
integration angle for a decaying DM signal also has a strong impact on
the ranking and on the error bars, more strongly than in the case of
DM annihilation. In particular, for $\alpha_{\rm int}=0.1^\circ$
(bottom panel), most targets have very similar $D$-factors and the
increased error bars make the ranking less obvious.

\paragraph*{Comparison to other works.} The availability of 
independently-derived $D$-factors for dSphs in the literature remains
limited, making comparison  less straightforward than in the
case of annihilation.

\begin{itemize}
\item Although not published in the \citetads{2011MNRAS.418.1526C}
  study which focused on $J$-factors only, the $D$-factors for the
  eight `classical' dSphs were also obtained from our original
  analysis setup. As in the case of annihilation, these values (green
  dots in figure~\ref{fig:D_results}) are systematically lower than
  that obtained by the present analysis and this is connected, as for
  $J$, to the choice of the light profile.

\item We also compare our results to those of
  \citetads{2015ApJ...801...74G}, noting again that the conservative
  choice made for the size of the DM halo in that study leads to a
  deficit in the $D$-factors compared to our values. This deficit is
  more apparent here than for annihilation as the outer regions of
  the DM halo contribute more to the $D$-factors than to the
  $J$-factors.

\item Following the claims of X-ray line detections in galaxy
  clusters, \citetads{2014PhRvD..90j3506M} looked for such a signal in
  the dSph galaxies available in XMM-Newton data. In the absence of
  a signal, these authors used the mass derived 
  from \citetads{2010MNRAS.406.1220W} \footnote{\citetads{2010MNRAS.406.1220W} solve the spherical
    Jeans equation coupled to a MCMC technique, using a similar
    approach to ours but different profile (mass, light, anisotropy)
    parametrisations, to provide a robust mass estimate of several
    dSph galaxies.} and \citetads{2015ApJ...801...74G}    
    to set constraints on a sterile neutrino DM
  scenario. In doing so, rather than writing the $D$-factor as given
  by equation~\ref{eq:J}, these authors use the point-like
  approximation and use instead $D^{\rm point}(\alpha_{\rm int}) = M_{\alpha_{\rm
    int}}/d^2$, where $d$ is the distance to the galaxy and
$M_{\alpha_{\rm int}}$ is the enclosed mass.\footnote{\citetads{2014PhRvD..90j3506M}
  do not explicitly call this
  quantity the $D$-factor; it simply appears as part of the overall
  flux definition. In the point
  like approximation, this mass is the mass enclosed in a sphere of
  radius $\alpha_{\rm int} \times d$. It is not the mass contained in
  the volume defined by the intersection of the line-of-sight cone and
  the dSph spherical halo.} \citetads{2010MNRAS.406.1220W} provide the
mass $M_{1/2}$ (and the corresponding error bars) contained within the
deprojected half-light radius $r_{1/2}$, corresponding to an
integration angle $\alpha_{1/2}=r_{1/2}/d$. This angle is closely
related to our definition of the optimal integration for decay
$\alpha^D_c=r_h/d$, where $r_h \approx 0.75 \times r_{1/2}$ is the
projected half-light radius.\footnote{\citetads{2010MNRAS.406.1220W}
  provides useful fitting formulae to relate the projected and
  deprojected half-light radii, for a variety of light profiles.} For
each dSph in the \citetads{2010MNRAS.406.1220W} sample, we compute
$D^{\rm point}(\alpha_{1/2})=M_{1/2}/d^2$ (orange triangles in
figure~\ref{fig:D_results}). Despite the fact that $\alpha_{1/2} >
\alpha^D_c$, our $D$-factors are generally higher than the ones
derived from the \citetads{2010MNRAS.406.1220W} data. Computing
$M_{1/2}$ from our MCMC chains provides, in general, values that are
compatible with those of \citetads{2010MNRAS.406.1220W}; this excludes
the mass reconstruction from being the sole origin of the differences.
The main remaining difference may lie in the point-like approximation, which
does not account properly for the full volume of the DM halo
being intercepted by the line-of-sight. While this may not be an
inappropriate assumption for the strongly peaked annihilation signal,
it results in a significant deficit for the $D$-factors.
\end{itemize}

\subsection{Discussion}
\label{subsec:target_by_target}

In the previous sub-sections, we presented a ranking of the Milky Way dSph
satellites as potential  targets for dark matter annihilation/decay surveys
based on the estimated values and uncertainties  of their $J$-factors or
$D$-factors. However, it is important to note that while our analysis has 
marginalised over many of the modelling uncertainties, issues such as the
dynamical status of individual dSphs and evidence pointing to cored profiles
in a number of the `classical' dSphs are not accounted for.

First, it is possible that some of the dSphs, especially the `ultrafaints',
are not currently in dynamical equilibrium. In particular, a number of
authors have presented evidence that the UMa2 dSph, which  occupies the top
position in both the $J$- and $D$-factor rankings, is currently experiencing
strong tidal disturbance by the Milky
Way~\citepads{2007MNRAS.375.1171F,2010AJ....140..138M,2013MNRAS.433.2529S}.
While it is not clear that  this has inflated the velocity dispersion of
UMa2, some caution is advisable before selecting UMa2
as a prime candidate for indirect dark matter detection surveys.

Secondly, the precise nature of some `ultrafaint' galaxies is still
uncertain---it is possible that some are more closely related to star
clusters and do not, in fact, contain dark matter. For example, the most
recent study of Wil1 describes it as "A Probable Dwarf Galaxy with an
Irregular Kinematic Distribution" and notes that foreground contamination
and unusual kinematics make the determination of its dark matter content
difficult~\citepads{2011AJ....142..128W}. Target selection for dark matter
surveys must take account of such uncertainty when weighting candidates for
study.

The Sextans, Ursa Minor and Draco `classical' dSphs feature in the top five
of both the $J$- and $D$-factor rankings. The  larger kinematic samples in
these objects make the Jeans modelling more robust (as indicated by their
smaller confidence  intervals in figures~\ref{fig:J_results}
and~\ref{fig:D_results}). However, in the case of
Sextans~\citepads{2013ApJ...777...65L,2004MNRAS.354L..66K}  and Ursa
Minor~\citepads{2014MNRAS.442.1718P,2005ApJ...630L.141K}, it has been
suggested that the presence of kinematic substructures indicate that the
dark matter halos of these objects are cored  rather than cusped. In
principle, this could be taken into account via priors on the halo slope
within our Bayesian analysis, but as shown in figure~15 of
\citetads{2011MNRAS.418.1526C}, this would not actually change the
conclusions for most integration angles (i.e., $\alpha_{\rm
int}\gtrsim 0.1^\circ$).


Thus, of the seven dSphs which are within the top five of the $J$ and $D$
rankings, additional evidence for 4 of them suggests that the results of
equilibrium dynamical modelling might not be sufficient to characterise
their suitability as targets for indirect dark matter detection. Clearly, it
is imperative that further data are obtained on potential candidates
before future surveys  select their target lists. New non-equilibrium
modelling approaches such as that presented in Ural et al. (2015, in press)
will also be important in placing the target selection on a secure footing.


\section{Conclusions}
\label{sec:conclusions}

Dwarf spheroidal galaxies have been widely targetted in for searches for
annihilating dark matter in the Galaxy. This has enabled $\gamma$-ray
telescopes to set very
stringent limits on the DM annihilation cross section which are now
beginning to impinge on the cosmologically preferred value for a
thermal relic. Reliable estimates of the dSph $J$-factors and
associated error budgets are clearly crucial in this regard. This is
especially true for `stacking' analyses that use data on several dSph
galaxies simultaneously to improve the sensitivity. 
Ranking of the dSphs, according to their $J$- (and $D$-) factors, is
also mandatory to optimise the strategy of pointed
observations. In case of a positive detection in a given target, this
will inform the strategy for subsequent observations aiming to
validate the DM hypothesis.

This study --- follow-up of our previous effort
\citepads{2011MNRAS.418.1526C} --- extends and improves the
reconstruction of the astrophysical factor for dSph galaxies in
several ways:
  \begin{itemize}
    \item We use the optimised analysis setup proposed in
      \citetads{2015MNRAS.446.3002B}: the parametrisation of the
      ingredients of the Jeans analysis are kept as general as
      possible to minimise biases. In the spirit of
      \citetads{2011MNRAS.418.1526C}, we adopt very weak priors to
      have as data-driven an analysis as possible.
    \item We rely on an improved analysis of light profiles
      (Walker et al., in prep.)  for better $J$-factor reconstruction
      \citepads{2015MNRAS.446.3002B}, and also include recent
      kinematic data (e.g. for Draco, see
      \citealtads{2015arXiv150302589W}) in the analysis.
    \item We test the impact of the choice of the likelihood function
      on the results: the performance and consistency of binned and
      unbinned analyses are validated on real and mock
      data. Furthermore, contamination from foreground stars and
      any associated impact on the $J$-factors are also investigated, using
      the membership probability  of stars when available.
    \item In addition to the 8 `classical' dSph galaxies in
      \citetads{2011MNRAS.418.1526C}, re-analysed here, the results
      for 13 `ultrafaint' dSphs are now also provided --- both the
      $J$-factors for annihilating DM and the $D$-factors for decaying
      candidates.
  \end{itemize}

The most important result of our study is the ranking (median and CIs)
of the astrophysical factors shown in figures~\ref{fig:J_results}
(annihilation) and \ref{fig:D_results} (decay), and summarised in
Table~\ref{tab:results}.  {\tt ASCII} files for the median and 68\%
and 95\% CIs for a large range of integration angles can be retrieved
from the Supporting Information submitted with this paper. Our findings can be summarised thus:
\begin{enumerate}
   \item The unbinned Jeans analysis of stars whose membership
     probability is $>0.95$ gives the most stringent constraints and is
     appropriate to deal conservatively with possible contaminations.
   \item Our $J$- and $D$-factors are in general consistent with other
     calculations, though several differences are observed: using a
     more flexible light profile (cf. the usually adopted Plummer
     profile) slightly increases the astrophysical factor for several
     dSphs; using a more flexible anisotropy profile (w.r.t. the
     usually assumed constant) slightly enlarges the $J$-factor CIs,
     providing more realistic uncertainties. This also mitigates
     possible biases \citepads{2015MNRAS.446.3002B}.
   \item Uncertainties on the astrophysical factors ($J$ and $D$) from
     this data-driven analysis are directly related to the sample size
     used for the analysis (large error bars for `ultrafaints', small
     error bars for `classicals'). We believe this better
     accommodates a possible non-universality in the properties of
     these objects.
   \item The ranking of the targets (according to their median values)
     slightly depends on the integration angle. At the optimal angle
     $\alpha_{c}^{J}\approx2r_h /d$, the `classical' dSphs UMi and
     Draco are confirmed as the potentially-brightest and most favoured targets
     in terms of $J$-factors.  The `ultrafaint' objects UMa~2 and
     Coma outrank them, but suffer from larger uncertainties, and in
     particular their lower 95\% CIs are lower than those of UMi or
     Draco. For decaying DM, Sextans appears as the brightest `classical'
     dSph at $\alpha_{c}^{D}\approx r_h /d$, while UMa~2 remains the
     brightest `ultrafaint' target. Not discussed here is the
     frequency-dependent astrophysical background that may affect this
     ranking. This would require an instrument-specific analysis that
     goes beyond the scope of this paper.
   \item The astrophysical factors of Segue~I might be highly uncertain due to probable stellar contamination and few kinematic data, and this object does not make it among the top ten targets (Bonnivard, Maurin \& Walker, in prep.). (Re)-analyses of membership probabilities for other `ultrafaints' would be helpful to probe the level of contamination in these objects.

\end{enumerate}

The 9 new potential dSphs discovered in the DES survey
\citepads{2015arXiv150302584T,2015arXiv150302079K} and already
searched for in Fermi-LAT data
\citepads{2015arXiv150302632T,2015arXiv150302320G} call for a
continued effort on this topic (see also \citealtads{2015ApJ...802L..18L,2015arXiv150306216M,2015arXiv150308268K}
for three other recently discovered dSphs). Careful studies of their astrophysical
factors will likely be difficult due to the small amount of kinematic
data expected for these objects. As underlined in this study, it is all the more important to explore the
limitations of kinematic analyses for such small stellar samples.  This
may prove crucial to understand whether the recently published limits
from 6 years of Fermi-LAT observations on 15 dSphs \citepads{2014arXiv1410.2242G,2015arXiv150302641F}
can be significantly improved or not, and/or if better targets exist
for the forthcoming Cherenkov Telescope Array \citepads{2011ExA....32..193A}.

\section*{Acknowledgements}

This work has been supported by the ``Investissements d'avenir, Labex
ENIGMASS", and by the French ANR, Project DMAstro-LHC,
ANR-12-BS05-0006.  This study used the CC-IN2P3 computation centre of
Lyon. JIR would like to acknowledge support from SNF grant
PP00SectionP2\_128540/1. SS thanks the DNRF for the award of a Niels
Bohr Professorship. MGW is supported by National Science Foundation grants
AST-1313045 and AST-1412999.

\appendix


\section{Unbinned analysis - validation on mock data}
\label{app:unbinned}

\citetads{2015MNRAS.446.3002B} thoroughly tested the Jeans analysis
using binned velocity dispersion profiles (see
equation~\ref{eq:likelihood_binned}), but unbinned analyses
(equation~\ref{eq:likelihood_unbinned}) are also often used in the
literature
(e.g. \citealtads{2008ApJ...678..614S,2009JCAP...06..014M,2015ApJ...801...74G}).
Here, we conduct a careful comparison using a large set of mock data
to determine the merits and limits of each approach, and select the
optimal setup for our analysis.

\paragraph*{Mock data set}
To this purpose, we employed the same suite of mock data set used
previously by
\citetads{2011ApJ...733L..46W,2011MNRAS.418.1526C,2015MNRAS.446.3002B}. It
consists of 64 models covering a large variety of DM density profiles
(from cored to NFW-like cuspy profiles), stellar light profiles and
velocity anisotropy values (between $\beta_\text{ani} = -0.45$ and
$\beta_\text{ani} = +0.3$, with constant anisotropy). We refer the
reader to the papers quoted above for a more complete description of
this mock data set. For each model, we draw samples of $N = 30$
(small), $1000$ (medium) and $10000$ (large) stars, mimicking
`ultrafaint', `classical' and `ideally observed' dSphs. No foreground
contamination is added to the data sets, and all the objects are fixed
to a distance $d = 100$ kpc.

\paragraph*{Analysis setup}
For each 64 mock models and sample sizes combination, we run the Jeans
analysis using either the binned\footnote{Note that we did not take
  into account the radius uncertainty on the velocity dispersion
  profiles ($\Delta R_i$ in equation~\ref{eq:sigma_tot}) in the
  \citetads{2015MNRAS.446.3002B} analysis.} or unbinned likelihood
functions described in Section \ref{subsec:Likelihood} (see
Eqs.~\ref{eq:likelihood_binned} and \ref{eq:likelihood_unbinned}
respectively). We fix the light and anisotropy parameters to their
true values (i.e. the ones used to generate the mock data), in order
to disentangle the effects of the likelihood functions to that
originating from the physical parameters. For each model we then
compute the $J$- and $D$-factors as a function of the integration angle
$\alpha_{\rm int}$.

\begin{figure}
\includegraphics[width=1.\linewidth]{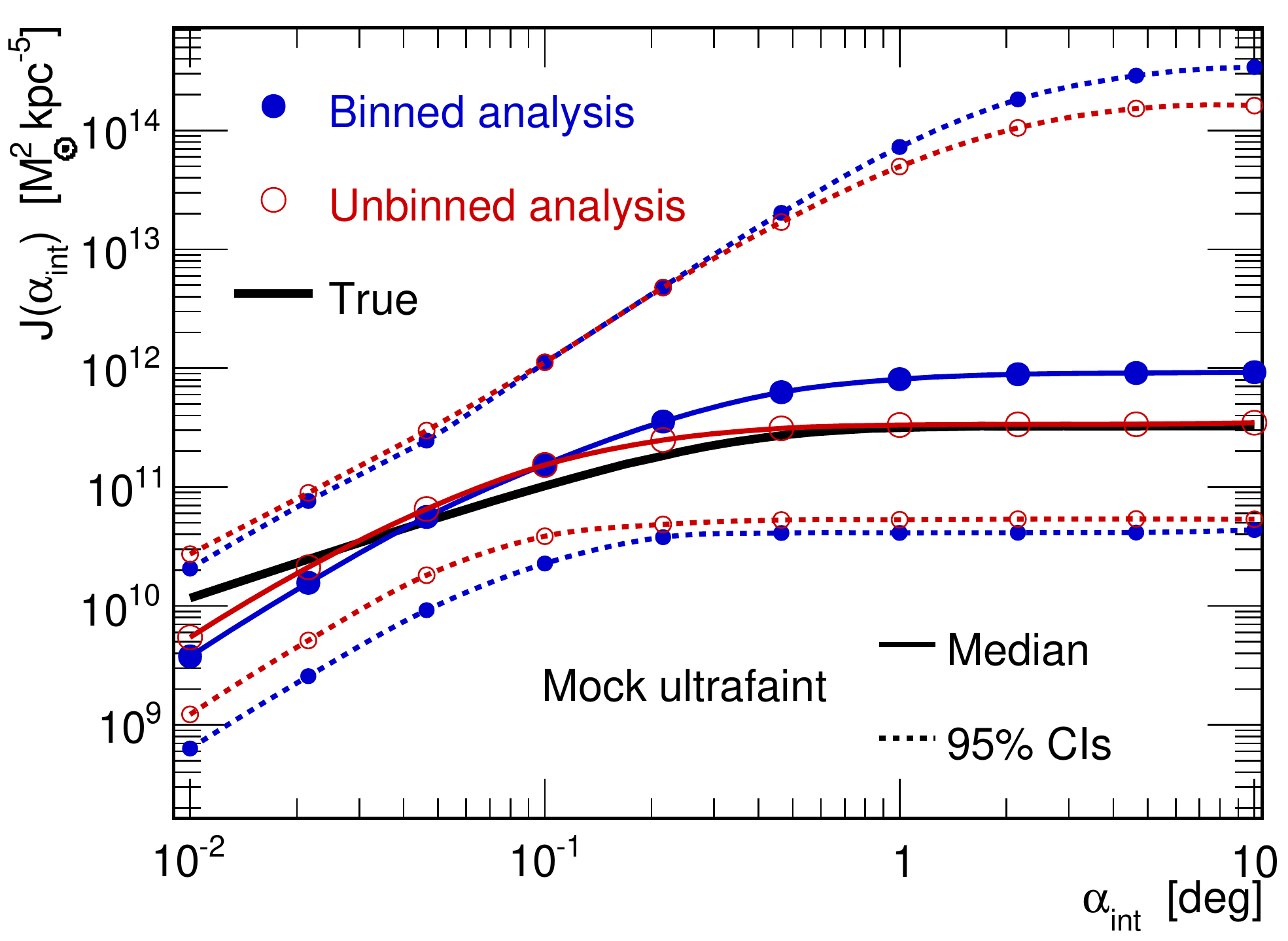}
\caption{Median value (solid) and 95\% CIs (dashed) of the $J$-factors
  as a function of the integration angle $\alpha_{\rm int}$,
  reconstructed using either a binned (blue filled circles) or an
  unbinned (red empty circles) Jeans analysis on a mock `ultrafaint'
  dSph. Both analyses lead to very similar results, with the true
  $J$-factor (black) being encompassed within the reconstructed CIs.}
\label{fig:J_Unbinned_Mock}
\end{figure}

\paragraph*{Effects of unbinned analysis}
First, we find that using either the binned or the unbinned likelihood
function leads to very similar results for the 64 models, regardless
of the sample size. This is illustrated in figure
\ref{fig:J_Unbinned_Mock}, by comparing the $J$-factors obtained using
either the binned (blue filled circles) or the unbinned (red empty
circles) analysis to the true value (solid back line, computed using
the DM parameters used to generate the mock data), for a typical
`ultrafaint'-like mock dSph.

\begin{figure}
\includegraphics[width=1.\linewidth]{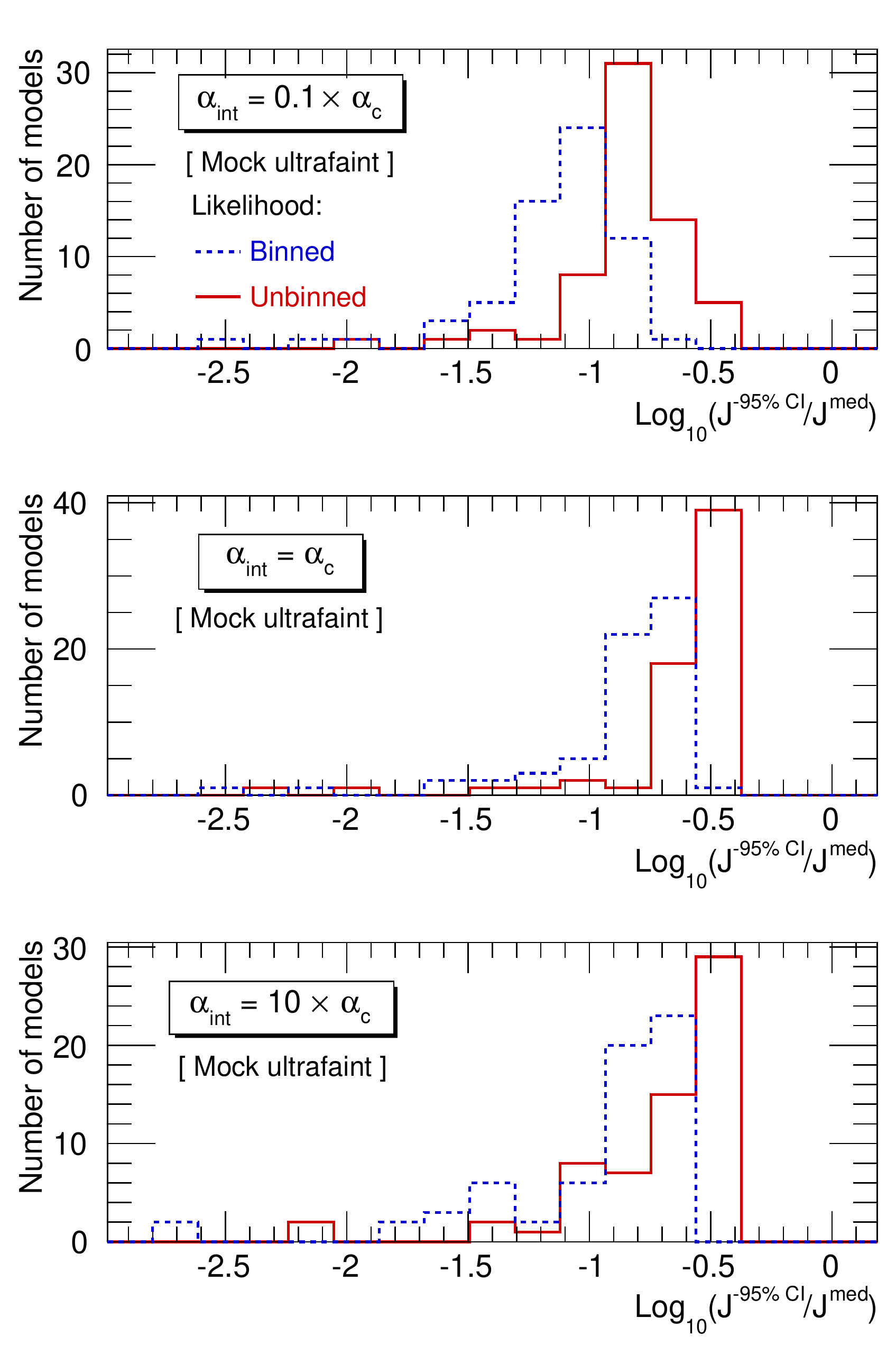}
\caption{Distributions of $J^{-95\% \rm CI}/J^{\rm med}$, obtained for
  the 64 mock `ultrafaint' dSphs using either the binned (dashed blue)
  or the unbinned (solid red) analysis. For the three integration
  angles considered ($\alpha_{\rm int} = 0.1 \times \alpha_c$,
  $\alpha_c$ and $10 \times \alpha_c$, from top to bottom), the
  unbinned analysis allows to reduce the mean $J^{-95\% \rm CI}/J^{\rm
    med}$ by a factor two. The effect is much less pronounced for
  larger data sets.}
\label{fig:J_Unbinned_Mock_Histo}
\end{figure}
The main difference between the two analyses lies in the CIs, as the
unbinned analysis is found to be more constraining than the binned
approach for `ultrafaint-like' dSphs. We show in figure
\ref{fig:J_Unbinned_Mock_Histo} the distributions of $J^{-95\% \rm
  CI}/J^{\rm med}$, i.e. the ratio of the lower 95\% CI to the median
$J$-factor, obtained for the 64 mock `ultrafaint' dSphs using either
the binned (dashed blue) or the unbinned (solid red) analysis. The
integration angle is set to $0.1 \times \alpha_c$ (top), $\alpha_c$
(middle) and $10 \times \alpha_c$ (bottom). For each integration
angle, the mean value of $J^{-95\% \rm CI}/J^{\rm med}$ is reduced by
a factor two for the mock `ultrafaint' dSphs when using the unbinned
analysis. The effect is much less pronounced for the medium
(`classical'-like) and large data samples. We note a similar effect on
the upper 95 \% CIs (not shown). For the $D$-factors, the effect is
also present for the `ultrafaint'-like dSphs, but less so, with a 
reduction of the CIs by $\sim 30 \%$ when using the unbinned analysis.

Finally, we have checked that no bias is introduced when using the
unbinned analysis on the 64 mock models. We therefore advocate the use
of the unbinned analysis when dealing with small data samples, as it
allows a significant reduction of the statistical uncertainties.

\section{Size of the DM halo}
\label{app:rvir}
As pointed out in Section \ref{subsec:size}, the way one defines the
DM halo size could influence the values of the astrophysical
factors. In order to quantify this effect, we performed several tests
on the mock dataset described in Appendix~\ref{app:unbinned}.

\begin{figure}
\includegraphics[width=1.\linewidth]{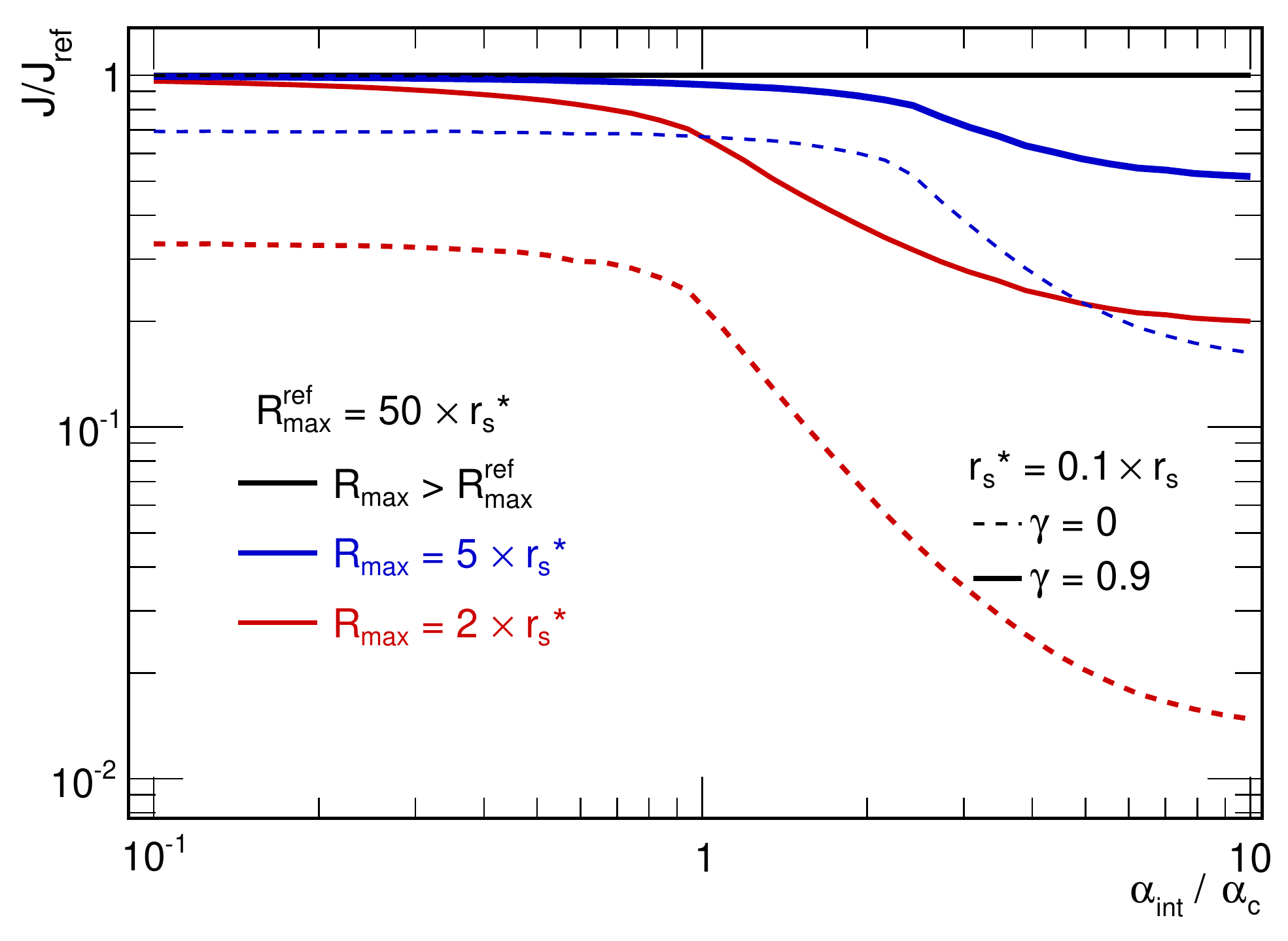}
\caption{Ratios of $J$-factors obtained with three different halo
  sizes $R_\text{max}$ (in black, blue and red colours) to a reference
  $J$-factor with $R_\text{max}^{\rm ref} = 50 \times r_{s}^{*}$. The
  smaller is $R_\text{max}/R_\text{max}^{\rm ref}$, the more
  underestimated is the $J$-factor. The effect is stronger for core
  (dashed lines) than for cuspy DM profiles (continuous lines).}
\label{fig:J_Jref}
\end{figure}

\paragraph*{Impact of halo size on $J$- and $D$-factors: tests on mock data}
First, using two typical mock models (with either cuspy or core DM
density profile), and for three different choices of halo maximum
radius $R_\text{max}$, we compute the $J$-factor as a function of the
integration angle $\alpha_{\rm int}$. Figure \ref{fig:J_Jref} compares
the results to a reference value obtained with $R_\text{max}^{\rm ref}
= 50 \times r_{s}^{*}$, chosen arbitrarily. Unsurprisingly it shows
that an underestimation of the halo size leads to an underestimation
of the $J$-factor, the effect being stronger for the core than for the
cuspy DM profile. The $J$-factor can be underestimated by up to 70\%
at the critical angle $\alpha_c$ for a core profile and a halo size
strongly underestimated (factor 25 too small). The effect is even more
important for the $D$-factors (underestimation by $\sim80\%$ at
$\alpha_c$ for the same model mentioned before, not shown).

\begin{figure}
\includegraphics[width=1.\linewidth]{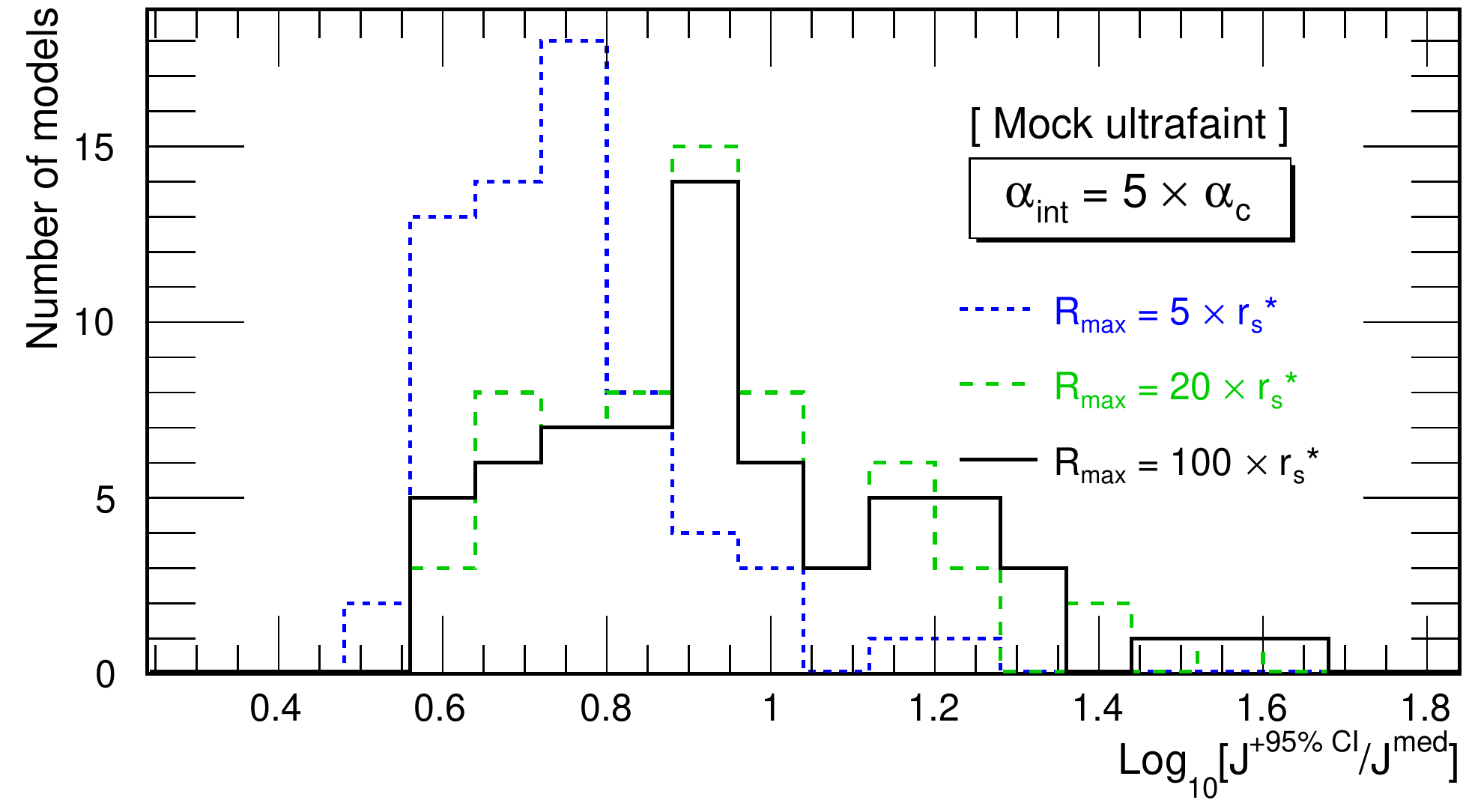}
\caption{Distributions of $J^{+95\% \rm CI}/J^{\rm med}$ obtained for
  the 64 mock `ultrafaint' dSphs at $\alpha_{\rm int} = 5\times
  \alpha_c$.  Three halo sizes are considered (short dashed blue,
  $R_\text{max} = 5 \times r_{s}^{*}$; long dashed green, $R_\text{max}
  = 20 \times r_{s}^{*}$; solid black, $R_\text{max} = 100 \times
  r_{s}^{*}$). At such large integration angles, smaller halo sizes
  reduce the credibility intervals. Underestimation of the halo size
  can then lead to underestimation of the CIs.}
\label{fig:Histo_CIs_Rvir}
\end{figure}

To evaluate the response of the CIs of the $J$-factors to the halo
size, we run the Jeans analysis on the entire set of mock dSphs (see
Appendix \ref{app:unbinned}). For each mock dSph, we fix three values
of $R_\text{max}$, and compute the $J$-factors and their CIs from the
reconstructed DM density profiles. We show in figure
\ref{fig:Histo_CIs_Rvir} the distributions of $J^{+95\% \rm CI}/J^{\rm
  med}$ for the three halo sizes, i.e. the ratio of the upper 95\% CI
to the median value, obtained for the 64 mock `ultrafaint' dSphs at
$\alpha_{\rm int} = 5\times \alpha_c$. For this large integration
angle, the credibility intervals shrink when the halo size gets
smaller, so that an underestimation of the halo size will also lead to
an underestimation of the credibility intervals. For the $J$-factors,
this effect appears only at large integration angles ($\alpha_{\rm
  int} > \alpha_c$).  For the $D$-factors, the effect is more
pronounced, and appears at all integration angles (not shown).

\begin{figure}
\includegraphics[width=1.\linewidth]{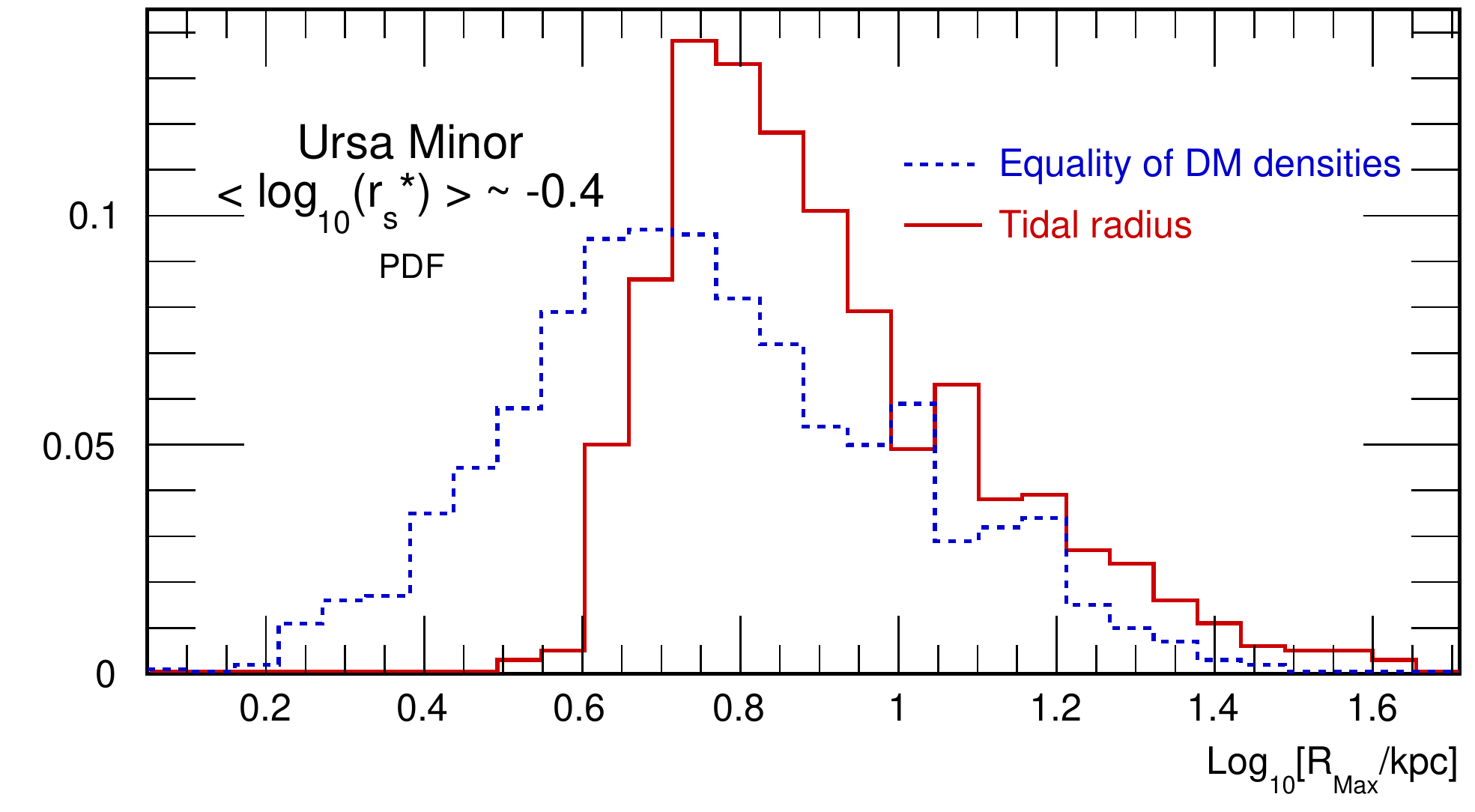}
\includegraphics[width=1.\linewidth]{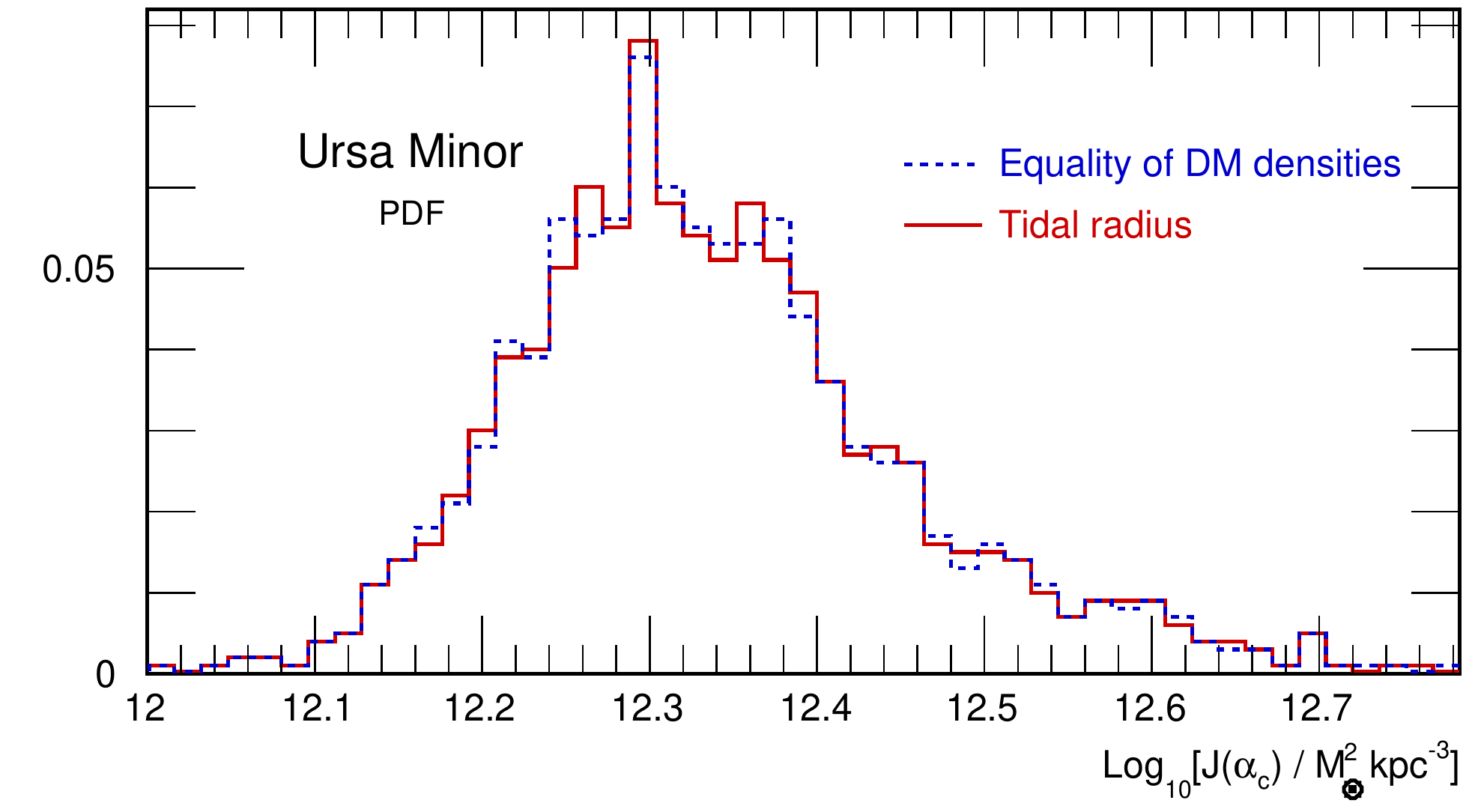}
\caption{\emph{Top:} Distribution of halo sizes obtained for Ursa
    Minor, using either the tidal radius (solid red) or the DM
    densities equality (dashed blue) estimation. Both distributions
    show a large spread over one order of magnitude, with the mean of the tidal
    radii distribution systematically larger than the mean estimation with
    the DM densities. Note that the mean light scale radius $\langle
    r_{s}^{*} \rangle$ (obtained by fitting the surface brightness
    data) is typically one order of magnitude lower than halo
    sizes derived here. \emph{Bottom:} Same as upper panel but for the
    $J$-factors. The systematic shift noted in the halo size does not
    propagate to the astrophysical factors.}
\label{fig:Pdf_Rvir_Umi}
\end{figure}

\paragraph*{Halo size estimation: comparison of the two methods}
In this work, we have considered two methods to estimate the halo
size, detailed in Section \ref{subsec:size}. The first method uses the
tidal radius as an estimation of the physical size of the halo; in the
second approach, it is evaluated as the radius where the halo DM
density and the MW DM density are equal. For a given dSph galaxy, the
halo size is computed for each DM model accepted by the MCMC
analysis. Figure \ref{fig:Pdf_Rvir_Umi} shows the distributions of
$R_\text{max}$ values obtained for Ursa Minor, using either the tidal
radius estimation (red solid) or the equality of DM densities (dashed
blue). Both distributions spread over more than one order of
magnitude, with the mean of tidal radii distribution being
systematically larger than the mean estimation from the DM density
equality. This behaviour is found in all the dSphs. This trend is
however not reflected on the astrophysical factors, for which both
methods give very similar results as shown in the bottom panel in
figure~\ref{fig:Pdf_Rvir_Umi}.


\section{Impact of contamination for `classical'-like dSph galaxies}
\label{app:pm}

Samples of stars from dSph galaxies may be contaminated by Milky Way
and/or stream interlopers. Contamination in the context of Segue~I will be discussed in Bonnivard, Maurin \& Walker (in prep.). The less spectacular case of Fornax is
presented below.

\begin{figure}
\includegraphics[width=\columnwidth]{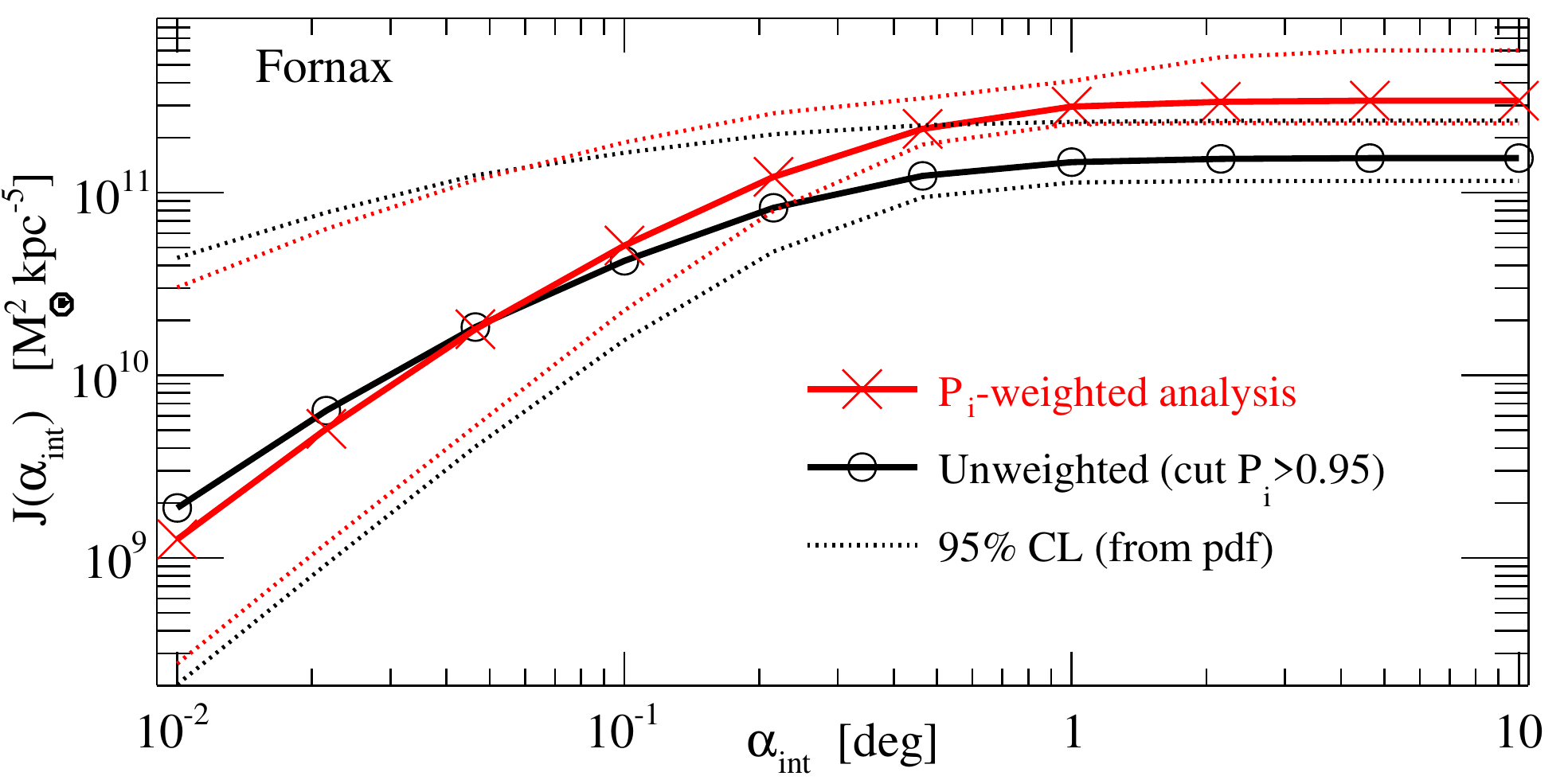}
\includegraphics[width=\columnwidth]{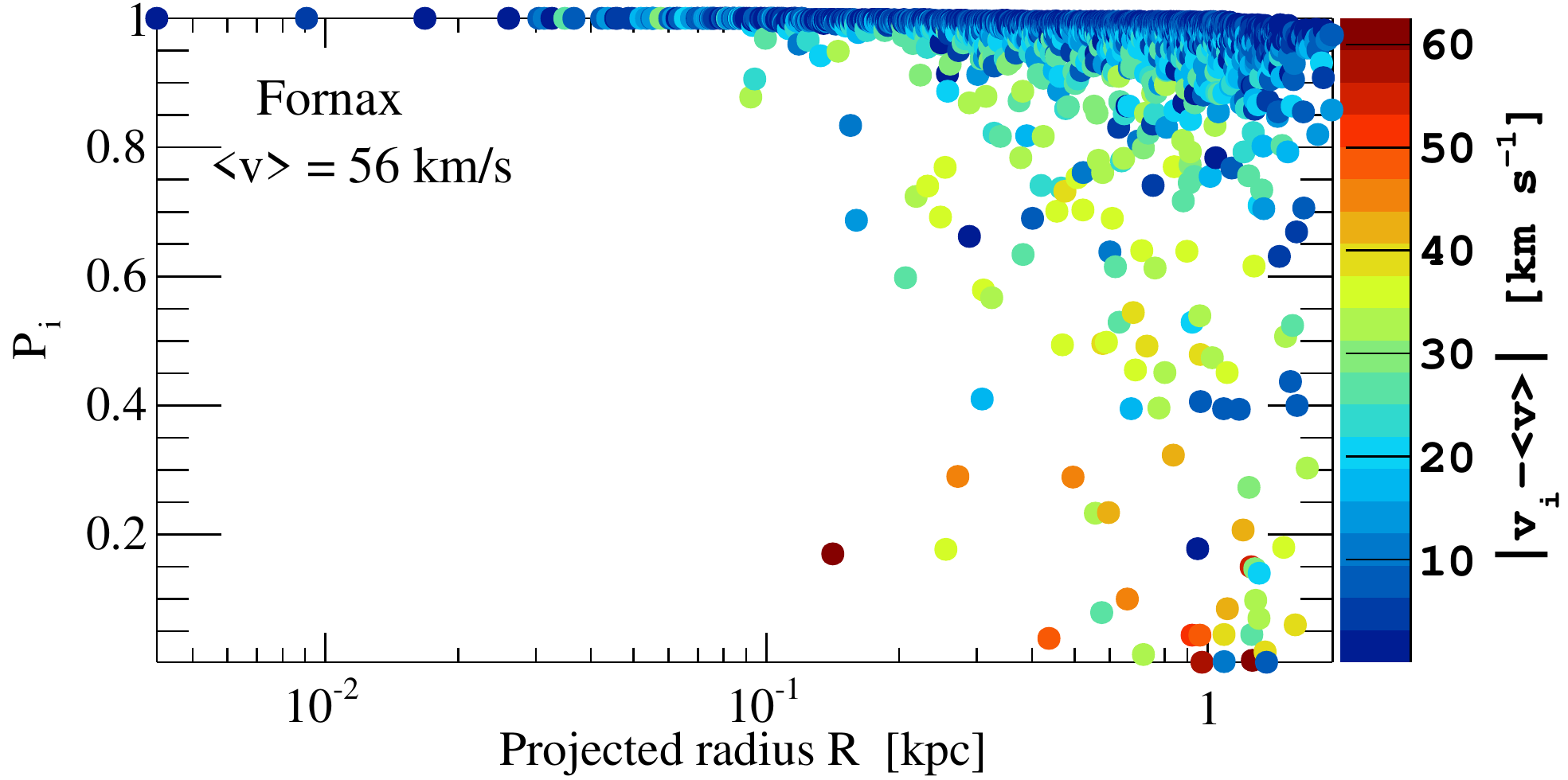}
\includegraphics[width=\columnwidth]{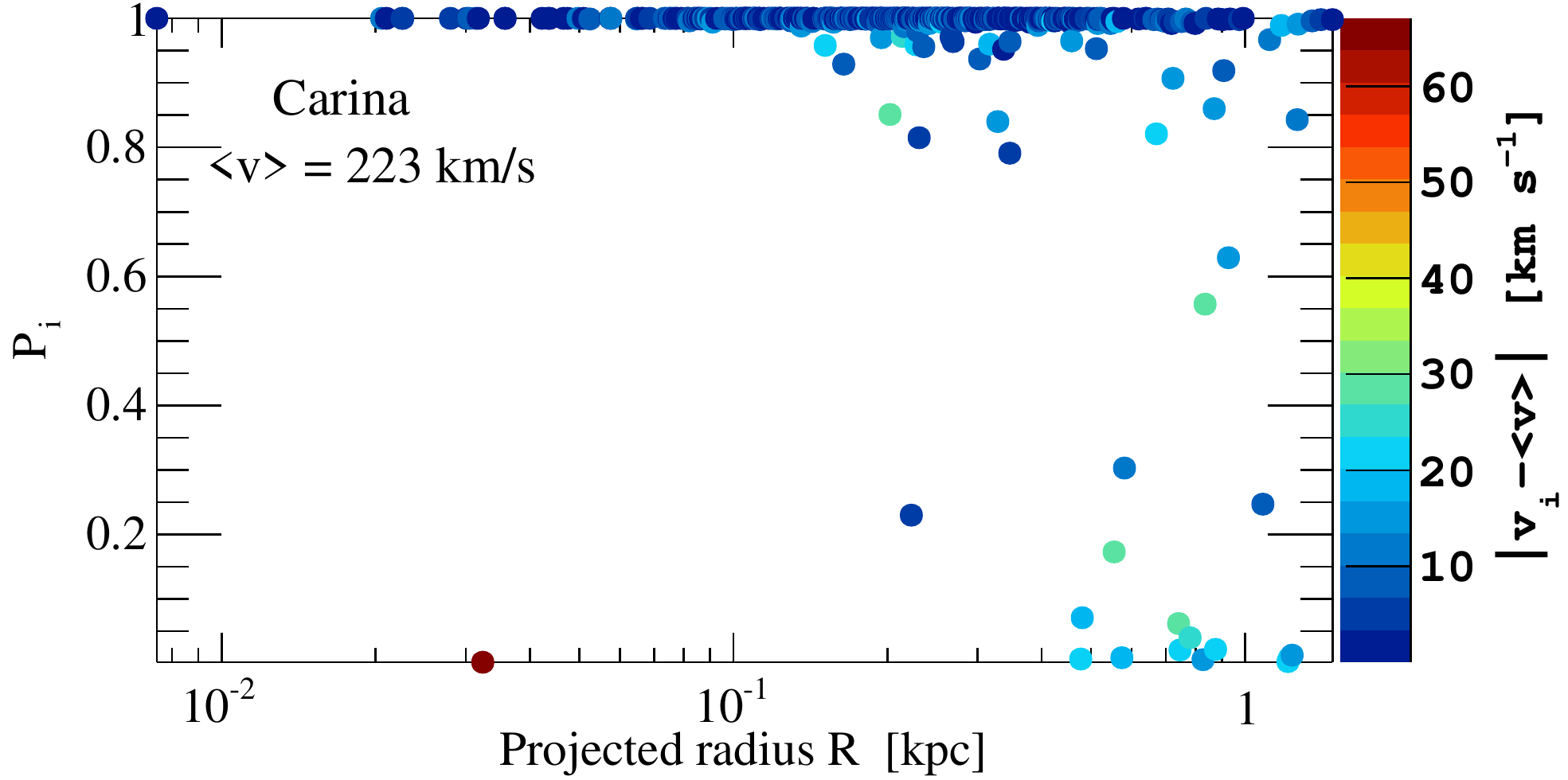}
\caption{Top panel: $J$-factor as a function of the integration angle
  $\alpha_{\text{int}}$ for the `classical' dSph Fornax, for an
  analysis with and without weighting by membership probabilities (see
  text). The lower panels show the membership probability $P_i$ of
  Fornax and Carina stars as a function of the projected radius
  $R$. The colour scale indicates the departure (red) from the mean
  velocity $\langle v\rangle$ of all stars (blue).}
\label{fig:J_Fornax_Membership}
\end{figure}

\paragraph*{Low impact contamination in Fornax}
Figure~\ref{fig:J_Fornax_Membership} shows the $J$-factors for Fornax
from both the $P_i$-weighted (red cross) and $P_i$-unweighted (cut
$P_i > 0.95$, black circles) analyses.  Contrarily to the other
`classical' dSphs, the two analyses give results which are in slight
disagreement at the two-sigmas level (especially at large radii). The
lower panels of figure~\ref{fig:J_Fornax_Membership} show the membership
probabilities as a function of the projected radius $R$ for Fornax
(middle panel) and Carina (bottom panel): 7 out of the 8 `classical'
dSph data display similar properties as Carina, i.e. most of the stars
have $P_i\gtrsim0.95$ with velocities close to the mean value $\langle
v\rangle$ for the object.  Fornax shows a significant fraction of
stars with intermediate $P_i$ values (especially at large radii),
whose velocity departs from the average. This is at the origin of the
difference between the two reconstructed $J$-factors (top panel).
Segue~I shows a much stronger dependence on the type of analysis, as will be discussed
in Bonnivard, Maurin \& Walker (in prep.).

\paragraph*{Impact of contamination checked on mock data}
Jeans analyses on mock data with controlled levels of contaminant are
helpful to investigate the impact and/or robustness of the
reconstructed $J$-factors. Among the contaminant-free mock data
described in \citetads{2015MNRAS.446.3002B}, we selected a couple of
models (with rising, flat, or decreasing velocity dispersion, from a
core and cusp DM profile) mixed with MW and stream stars (details will be presented in Bonnivard, Maurin \& Walker in prep.). For each model, we created 1000 data sets with
different levels of contamination from the MW and from the stream
(denoted $f_{\rm MW}$ and $f_{\rm stream}$ respectively, with $f_{\rm
  MW}+f_{\rm stream}\leq1$). Two sample sizes were drawn to mimic
`classical' ($\sim300-3000$ stars) and `ultrafaint' ($\sim30-100$
stars) dSph galaxies. After reconstructing membership probabilities
with the EM algorithm \citepads{2009AJ....137.3109W}, we selected the
contaminated mock data that have at least 10\% of their stars with
intermediate membership probabilities (0.1 $\leq P_i \leq$ 0.95). The
other sets of contaminated data correspond to a sample whose $P_i$ are
reconstructed with high certainty (similar to no contamination
case). We then ran the Jeans analysis, fixing the light and anisotropy
profiles parameters to their true values to factor out non-essential
ingredients of the analysis.

\begin{figure}
\includegraphics[width=\columnwidth]{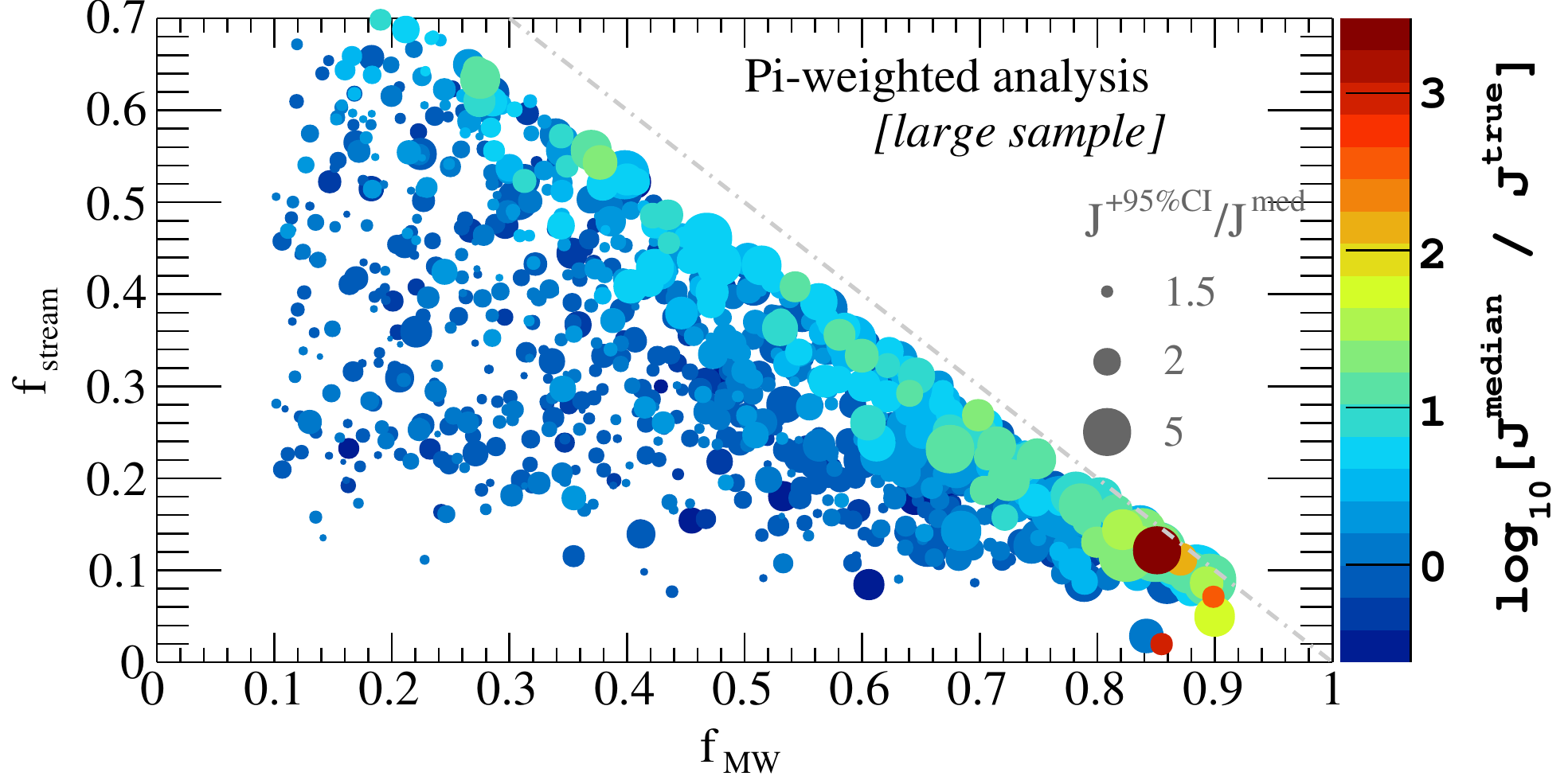}
\includegraphics[width=\columnwidth]{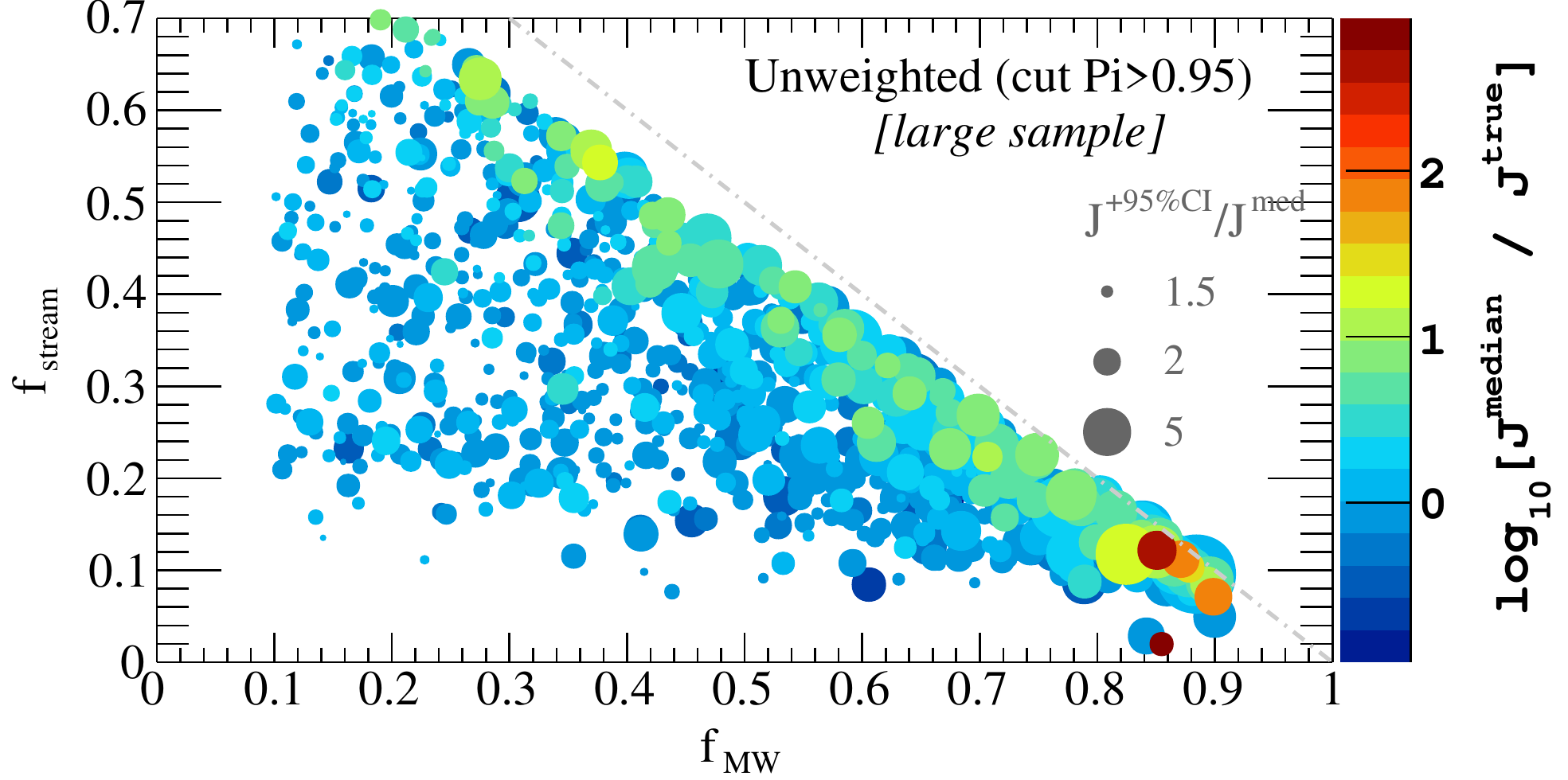}
\caption{Analysis of reconstructed $J$-factors ($\alpha_{\rm int} =
  \alpha_c\approx2r_h /d$, large sample size) of contaminated mock
  data. The $x$-axis (resp. $y$-axis) is the level of contamination
  from the Milky-Way (resp. stream) in the sample.  The colour scale
  shows the ratio of the reconstructed to the true $J$ value.  The
  size of the symbols correspond to the size of the CI on the
  reconstructed $J$-factor. See text for the discussion.}
\label{fig:J_weighted_vs_unweighted_large}
\end{figure}

The results for the small sample sizes, in the
context of Segue~I's analysis, will be presented in Bonnivard, Maurin \& Walker (in prep.). Here, we present the result for the larger sample
sizes in relation to Fornax, and show that $J$-factors for such samples
are much less affected by
contamination. Figure~\ref{fig:J_weighted_vs_unweighted_large} shows a
comparison of the reconstructed to the true value for both the
$P_i$-weighted and $P_i$-unweighted (cut $P_i > 0.95$) Jeans analyses
on adversely contaminated mock data: the former analysis relies on the
likelihood function equation~(\ref{eq:likelihood_unbinned_weights}), while
the second takes care of removing all stars with $P_i \leq 0.95$.  The
maximal departure from the true value is observed close to maximum
contamination (dashed line is $f_{\rm MW}+f_{\rm stream}=1$). However,
even in this worst case, the overshoot is within a factor 10 for most
of the models. Four models show more important overshooting (up to a
factor $10^3$), which are caused by a few misidentified contaminant
stars with both large departure from the mean velocity (up to 100
km/s) and large membership probability. These stars could be easily
identified and removed from a real data sample. We checked that none
of the `classical' dSph we studied had such extreme outliers.

This leads to the conclusion that objects with a significant number of
stars, as a result of having their membership probabilities robustly
recovered \citepads{2009AJ....137.3109W}, have a robustly
reconstructed $J$-factor, even in the presence of large
contamination. Regardless of the sample size, another conclusion is
that whenever the fraction of stars with intermediate membership
probabilities (0.1 $\leq P_i \leq$ 0.95) becomes significantly
different from zero, the $P_i-R-|\Delta v|$ plot is useful to identify
likely contaminated objects: in the case of `ultrafaint' dSph
galaxies, the impact on the $J$-factor can be very important (Bonnivard, Maurin \& Walker in prep.).

\label{lastpage}
\bibliography{dsphs_data}
\end{document}